%% file: Main.tex
\documentclass[superscriptaddress,pre,aps]{revtex4-2}

\usepackage[pdfstartview=FitH,
  breaklinks=true,
  bookmarksopen=false,
  bookmarksnumbered=true,
  colorlinks=true,
  linkcolor=black, 
  citecolor=black,
  urlcolor=black,
  pdftitle={},
  pdfsubject={}
  ]{hyperref}

\usepackage{graphicx}

\usepackage{xcolor}
\definecolor{amethyst}{rgb}{0.54, 0.17, 0.89}
\definecolor{coral}{rgb}{1.0, 0.3, 0.4}

\usepackage{amssymb}
\usepackage{amsmath}
\usepackage{epsfig}
\usepackage{chemarr}
\usepackage{url}
\usepackage{natbib}
\usepackage{color}
\usepackage{bm}
\usepackage{comment}
\newcommand{\arctanh}{\mathop{\mathrm{arctanh}}\nolimits}

\usepackage[capitalize]{cleveref}   

\begin{document}

\title{Stochastic Thermodynamics of Associative Memory}
\author{Spencer Rooke}
\affiliation{David Rittenhouse Laboratory, University of Pennsylvania, Philadelphia, PA 19104, USA}
\affiliation{Computational Neuroscience Initiative, University of Pennsylvania, Philadelphia, PA 19104, USA}
\author{Dmitry Krotov}
\affiliation{IBM Research, Cambridge, MA, USA}
\author{Vijay Balasubramanian}\thanks{Equal contribution}
\affiliation{David Rittenhouse Laboratory, University of Pennsylvania, Philadelphia, PA 19104, USA}
\affiliation{Computational Neuroscience Initiative, University of Pennsylvania, Philadelphia, PA 19104, USA}
\affiliation{Santa Fe Institute,
1399 Hyde Park Road, Santa Fe, NM 87501, USA} 
\author{David Wolpert}\thanks{Equal contribution}
\affiliation{Santa Fe Institute,
1399 Hyde Park Road, Santa Fe, NM 87501, USA} 
\begin{abstract}
    Dense Associative Memory networks (DenseAMs) unify several popular paradigms in Artificial Intelligence (AI), such as Hopfield Networks, transformers, and diffusion models, while casting their computational properties into the language of dynamical systems and energy landscapes. This formulation provides a natural setting for studying thermodynamics and  computation in neural systems, because DenseAMs are simultaneously simple enough to admit analytic treatment and rich enough to implement nontrivial computational function. Aspects of these networks have been studied at equilibrium and at zero temperature, but the thermodynamic costs associated with their operation out of equilibrium are largely unexplored.  Here, we study the thermodynamic entropy production associated with the operation of such networks, and study polynomial DenseAMs at intermediate memory load. At large system sizes and intermediate and low load, we use mean field theory to characterize out-of-equilibrium properties, work requirements, and memory transition times when driving the system with corrupted memories. We characterize a failure mode of higher order networks not observed at zero temperature. Further, we develop a method for calculating work and power costs in the mean field limit. Finally, we find tradeoffs between entropy production, memory retrieval accuracy, and operation speed. 
\end{abstract}

\maketitle
\section{Introduction}

Models of neural computation inspired by interacting spin systems have a long history, and were famously popularized by Hopfield Networks and Boltzmann Machines \cite{Little1,Hopfield1,Boltzmann1}. In these networks, memory and computation are  governed by  energy landscapes, connecting neural dynamics to statistical mechanics.  Conversely, most modern Artifical Neural Network (ANN) architectures are designed without attention to energetic  landscapes governing dynamics. Thus, while modern ANNs  achieve remarkable performance on a wide array of tasks \cite{Perf1,Perf2,perf3,perf4}, the thermodynamic costs they incur are equally immense, especially when compared against neural networks found in nature which appear to have architectural and information coding adaptations to reduce metabolic cost \cite{levy1996energy,balasubramanian2001metabolically,attwell2001energy,balasubramanian2002test,perge2012axons,balasubramanian2015heterogeneity,levy2021communication,Comp1,Comp2,Comp3}.  Here we revisit classical energy-based models to derive theoretical insights for efficient network operation and design, with the goal of better understanding the thermodynamic footprint of computation by artificial networks.

We will employ the lens of stochastic thermodynamics, which provides a framework for describing non-equilibrium behavior and energetics of systems evolving under the influence of noise, an  approach which has been useful for characterizing driven systems in contact with thermal environments \cite{Thermo_Info,MolecularMachines}. The application of stochastic thermodynamics to information processing systems is  growing rapidly \cite{wolpert2024stochastic,LearningThermo} because neural network computation is out of equilibrium, driven, and often stochastic. Further, foundational work has established universal bounds on the thermodynamic costs of learning \cite{LearningThermo,Referee2} and information processing in multipartite systems \cite{Referee3,Referee4}. 
However, efforts in these directions have thus far largely focused on systems with few components and low dimensional state spaces \cite{Thermo_Info,Wolpert_2019,Barato_2015}. Instead, we study the thermodynamic cost of large networks implementing associative memory. 

\begin{figure}
    \centering
    \includegraphics[width=.90\textwidth]{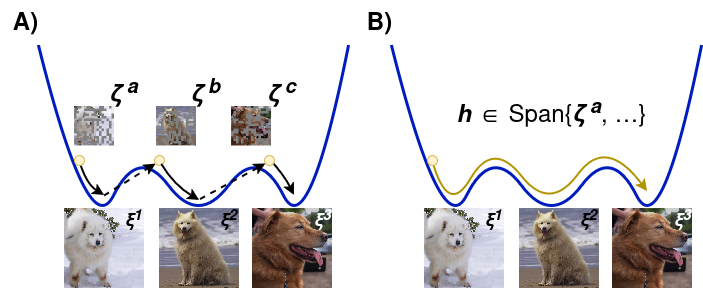}
    \caption{Memories ($\bm{\xi}^\mu$) are stored as energy minimizing network configurations in an energy landscape. We consider two modes of operation: \textbf{(A)} 
    We initialize the network in a partial memory ($\bm{\zeta}$), let it relax under Glauber dynamics, then do work to reinitialize the network into the next partial memory; \textbf{(B)} We do direct continuous work on the system through the control fields ${\bf h}$. We restrict ${\bf h}$ to be a linear combination of  corrupted memories. 
    }
    \label{fig:landscape}
\end{figure}

We focus  on  networks  such as Hopfield Networks and Dense Associative Memory Networks (DenseAMs) \cite{Hopfield1,Gardner,patternRecognition} implemented by interacting spins  modeling two-state neurons. Such networks are designed to recall a set of ``memories" from partial cues with minimal error. The desired recall can be achieved by preparing interactions between neurons such that system configurations associated with  memories are local energy minima and fixed points of the network dynamics. 
Usually, when initialized in some state, the network autonomously evolves to minimize its energy, eventually reaching a local minimum that corresponds to a stored memory (Fig.~\ref{fig:landscape}). Such networks can thus be utilized to correct corrupted versions of the patterns stored by a network \cite{krotov2025modern}. We will study such computational architectures at finite temperature, where gradients become stochastic, and free energies, rather than energies,  are minimized by dynamics.

A distinctive feature of DenseAMs is that they can store a much larger amount of information than conventional Hopfield Networks. A classical Hopfield Network with $N$ neurons can only store $\sim N$ generic memories \cite{Hopfield1,Sompolinsky2}. DenseAMs, on the other hand, can store a power law $\sim N^n$ ($n\geq2$ is a parameter of the energy function), or even an exponentially large $\sim \exp(\alpha N)$ number of memories \cite{patternRecognition, demircigil2017model}. Additionally, DenseAMs can be formulated in ways that resemble  useful structures  used in AI, such as convolutional layers \cite{krotov2021hierarchical}, attention layers \cite{allyouneed}, and transformer blocks \cite{hoover2023energy}. Furthermore, connections have been drawn between gradients of DenseAM energy landscapes and the minimization of score functions used in diffusion models \cite{Diffuse2, Diffuse1, pham2025memorization}.

DenseAMs also provide a powerful framework for discussing  information processing in biological neural networks. Multi neuron couplings, responsible for large information storage capacity, can be represented as effective theories for networks whose interactions are predominantly pairwise \cite{krotovlarge}. Astrocytes, which are non-neuronal cells in the brain, may provide a biological substrate for effective multineuronal couplings, similar to those that appear in DenseAMs \cite{kozachkov2025neuron}. Finally, simple models of sequential memory recall, which are similar to models of sequences of motor commands, have been designed using these ideas \cite{chaudhry2023long, karuvally2023general, herron2023robust}.  With these applications in mind, DenseAMs provide a natural setting for understanding thermodynamic costs in energy-based neural networks. In their simplest instantiations, DenseAMs implement  associative memory recall, do not require extensive training (patterns can be embedded in the energy landscape through Hebbian learning), and have natural interpretations in terms of energetics. The latter feature makes them  amenable to the tools of statistical mechanics.

While the equilibrium behaviour of Hopfield-like models is well understood \cite{Hopfield1,Gardner,patternRecognition,Sompolinsky1,Sompolinsky2}, the thermodynamic costs associated with such networks evolving to a stored pattern or driven by an external agent have remained largely unexplored, even though 
they are of great interest because biological and artificial neural networks often operate far from equilibrium.  
The operational costs can be understood thermodynamically in terms of the entropy produced during time evolution, reflecting irreversibility of  network dynamics. 
Entropy production in physical networks in turn leads to increased power consumption  and losses from heat dissipation. Understanding these costs may thus lead to insights for optimizing networks to minimize entropy production, and by extension, energy consumption and losses.

To this end, we explore the stochastic thermodynamics of DenseAM networks operating away from saturation (low to intermediate memory load, $\alpha=0$) and coupled to a single, fixed-temperature reservoir. Unlike the typical setting found in the literature, where the network evolves under greedy descent of an energy landscape, we consider networks evolving under a (temperature dependent) continuous time Markov process, in a thermodynamically consistent manner. In this setting, we characterize the dynamics of the network in a mean field limit, in terms of  alignment of the network state with each memory stored by the network.  While dynamic mean field theory (DMFT) has been applied previously to neural networks, to our knowledge we are presenting the first application of DMFT to calculate work and entropy production in  nonstationary, out-of-equilibrium neural network processes. We consider both the “typical” use of such networks, in which the state is initialized from a corrupted pattern, and the network’s response to rapid external driving that moves it through multiple memories. 

We present three main results: \textbf{(1)} We establish that dense networks with higher order nonlinearities have a failure mode of pattern completion at nonzero temperatures that is absent from lower order networks and when the temperature is zero; \textbf{(2)} We introduce a method that is exact in the mean field limit for calculating the amount of work expended in these networks under arbitrary fast driving; \textbf{(3)} We use this new method to demonstrate tradeoffs between entropy production, memory retrieval, and operation speed for a class of driving strategies.

We  start in Sec.~\ref{sec:background} by defining the network and its dynamics, along with standard terms used in non-equilibrium statistical mechanics.  In Sec.~\ref{sec:equilibrium} we characterize the equilibrium behavior of these networks at low to intermediate load. Next, in
Sec.~\ref{sec:dynamics} 
we consider relaxation dynamics at finite temperature, and characterize failure modes, relaxation times, recovery accuracy, and dissipation in polynomial networks of various order.   We then proceed to the general driven setting, and characterize work and power costs associated with driving the network with partial cues using coarse grained degrees of freedom. Importantly, the calculated costs is exact in the large system size limit, whereas typically these thermodynamic quantities are only well characterized in small systems. This allows us to compute work and power costs associated with arbitrary finite time control protocols. Using these tools, we show  that for a representative family of driving strategies, higher order networks require greater power during operation for equal operation speed. This demonstrates a tradeoff between performance and thermodynamic costs for networks of various orders. 
We conclude with a discussion in Sec.~\ref{sec:discussion}.

\section{Background}
\label{sec:background}

\subsection{Dense Associative Memory Networks}

We begin by considering networks of $N$ binary spins, $\sigma_i=\pm 1$,
with state space $\Omega$, and configurations $\bm{\sigma}\in \Omega$.  Suppose we also have a set $\{\bm{\xi}^\mu\}_{\mu=1}^p$ of $p$ memories. We want to store those memories as energy minima of the network that act as attractors of the  dynamics at low temperature.  The simplest Hamiltonian (energy function) that stores the memories is quadratic in the spins, with coupling matrix $J$ chosen as a sum of projections onto each memory. 
This yields the Hopfield model \cite{Hopfield1}:
\begin{align}
    \mathcal{H}_{\textrm{Hopf}}(\bm{\sigma}) = -\frac{1}{N} \bm{\sigma}^T J \bm{\sigma} -
    \bm{\sigma} 
    \cdot \sum_i \bm{h_i}(t)
    \;\;\; ; \;\;\;
    J_{ij} = \sum_{\mu=1}^p \bm{\xi}^\mu_i
    \bm{\xi}^\mu_j
    \label{eq:Hamil1}
\end{align}
Here, each $\bm{h}_i(t)$ represents a local field through which we can do work on the system.  In a realistic neural network setting, each $\bm{h}_i(t)$ may itself be comprised of a linear combination of neurons in an earlier network layer or represent sensory inputs to the network.   For simplicity, here we will assume there is a single local driving field.  Additionally, we will assume that each $\xi_i^\mu = \pm 1$ with equal probability and that the memories  are statistically uncorrelated. The coupling matrix $J$ constructed in (\ref{eq:Hamil1}) can store a  number of memories linear in the number of neurons, with critical capacity $p_C \sim 0.138 N$ at large $N$ \cite{Hopfield1,Sompolinsky1}. 
In fact, there are quadratic coupling matrices which lead to better storage capacity up to a maximum of $p_C\sim 2N$ \cite{Gardner_Interactions}. A more general family  of Hamiltonians, known as polynomial DenseAM networks, takes the form  \cite{patternRecognition}:
\begin{align}
    \mathcal{H}_{\textrm{DAN}}(\bm{\sigma}) = -\frac{1}{N^{k-1}}\sum_\mu (\mathbf{\bm{\sigma}}\cdot \bm{\xi}^\mu)^k
    - \bm{h}\cdot\bm{\sigma}
    \label{eq:Hamil2}
\end{align}
For $k=2$, this reproduces the Hopfield model. For $k>2$, the interactions are no longer pairwise, and involve multiple spins/neurons.
In terms of networks in the brain, these multi-neuronal interactions in (\ref{eq:Hamil2}) can arise from effective theories that omit descriptions of intermediate neurons. As we will be interested in the thermodynamics of such networks at large $N$, we have chosen a normalization that keeps the energy density extensive in the system size. With this normalization, the energy of a network when it is perfectly localized to a single memory is of order $N$. At zero temperature and large $N$, these networks can store $\alpha_k^* N^{k-1}$ memories,  where the capacity paramter $\alpha_k^*$ depends on the order of the nonlinearity and the allowed error at zero temperature \cite{Gardner,patternRecognition,StudentTeacher}. Thus the memory storage capacity increases rapidly with $k$. We will work with loads where the number of memories $p$ is below  saturation $p<<N^{k-1}$, as this simplifies our  analysis.

We will consider  these systems both in and out of equilibrium under continuous time Markovian dynamics:
\begin{align}
    \partial_t P(\bm{\sigma}) &=
    \sum_i \left[ \Gamma_i(S_i\bm{\sigma};t)P(S_i\bm{\sigma};t) 
    -\Gamma_i(\bm{\sigma};t)P(\bm{\sigma};t) \right] \label{eq:Master1}\\
    \Gamma_i(\bm{\sigma};t) &= 
    \frac{1}{2\tau}\bigg[1-\tanh(\frac{1}{2}\beta[\mathcal{H}(S_i\bm{\sigma};t)
        -\mathcal{H}(\bm{\sigma};t)])\bigg] \label{eq:Master2}\\
    S_i\bm{\sigma} &= (\sigma_1,...,-\sigma_i,...,\sigma_N)
\end{align}
Here, $P(\bm{\sigma})$ is the probability of a particular spin configuration $\bm{\sigma}$, transition rates associated to flipping spin $i$ are denoted $\Gamma_i$, the timescale of the dynamics is set by $\tau$, and $S_i$ acts on $\bm{\sigma}$ to flip spin $i$. While different choices for the transition rates $\Gamma_i$ are consistent with detailed balance at equilibrium, we have chosen the canonical transition rules for classical spin dynamics \cite{Glauber,IsingDynamics,GlauberNN,coolen}. 

We wish to utilize this network to perform simple computations in which the dynamics retrieves full patterns $\bm{\xi}$ from partial cues denoted $\bm{\zeta}$s.  We will use such dynamics to perform pattern matching such that the state of the system is driven to the nearest local energy minimum, which in encodes a particular memory (Fig.~1).

We formalize the process as follows: given a set of corrupted patterns $\{\bm{\zeta}^1, \bm{\zeta}^2...\}$, we want to recover the true memories represented by each. That is, we regard each $\bm{\zeta}^\nu$ as a fuzzy version of some stored memory $\bm{\xi}^\mu$, and the task is to recover an uncorrupted version of each memory. Under ideal operation, we want to do this quickly, accurately, and without doing too much work or generating too much heat. We will find that these three objectives are in tension under typical driving protocols. In standard use, we  initialize the network into a partial memory, let it relax, and then repeat.  Alternatively, we can drive the network by applying external fields $\bm{h}$. We assume that $\bm{h}$ only has information about partial memories, so we restrict ourselves to control strategies in which $\bm{h}\in\textrm{Span}\{\bm{\zeta}\}$.

\subsection{Stochastic Thermodynamics}
We will use the methods of stochastic thermodynamics,  a framework for describing systems evolving out of equilibrium while coupled to thermal environments, to characterize the networks described above with the dynamics in Eq.~\eqref{eq:Master1}. In this framework, thermodynamic quantities such as heat, work, and entropy production can be defined both along stochastic trajectories and at the level of ensembles \cite{Van_den_Broeck_2015,Van_den_Broeck_2010,MolecularMachines,Thermo_Info}. 

The first law must hold at both the trajectory and ensemble level. At the trajectory level, changes in the system energy $E = \mathcal{H}(\bm{\sigma}(t),t)$ can be decomposed into work done by external control parameters (associated with changes in the energy levels of the Hamiltonian), and heat exchanged with the environment. For our system, the control parameters are the external fields $\bm{h}$. At the ensemble level, changes in energy can be expressed:
\begin{align}
    d_t \langle E \rangle &=
    d_t \sum_{\{\bm{\sigma}\}}[\mathcal{H}(\bm{\sigma},t)P(\bm{\sigma},t)]
    =\dot{Q} + \dot{W}
\end{align}
where $d_t \equiv d/dt$ is the total derivative with respect to time. Rates of heat flow and work are thus identified from the chain rule:
\begin{align}
    \dot{Q} &= \sum_{\{\bm{\sigma}\}}[\mathcal{H}(\bm{\sigma},t)d_t P(\bm{\sigma},t)]
    \label{eq:HeatFlow}  \\
    \dot{W} &= \sum_{\{\bm{\sigma}\}}[d_t\mathcal{H}(\bm{\sigma},t) P(\bm{\sigma},t)] 
    = \langle d_t\mathcal{H}(\bm{\sigma},t) \rangle_{P(\bm{\sigma},t)}
    \label{eq:Work0}
\end{align}
Heat flows correspond to stochastic transitions between network states induced by the interaction with the thermal bath, while work corresponds to externally driven changes to the energy levels of states of the Hamiltonian, weighted probabilistically by state occupancy.
As heat is exchanged between the environment and the bath, and work is done on the system, entropy is produced simultaneously in the bath and the network. By the second law, the entropy produced in the joint bath+network system must be positive. The entropy of the system is simply proportional to the Shannon Entropy:
\begin{align}
    S_{\textrm{sys}}(t) = \langle -\ln[P(\bm{\sigma},t)]\rangle_{P(\bm{\sigma},t)}
\end{align}
where we have set $k_B=1$. The thermal bath is much larger than the system and assumed to be at equilibrium at all times. As a result, the only entropy produced in the bath is due to heat flowing between the bath and the network. 
The total (irreversible) entropy production in the bath+network system is then:
\begin{align}
    \dot{S}_{\textrm{tot}} = \dot{S}_{\textrm{sys}}-\frac{1}{T}\dot{Q}\geq 0
    \label{eq:EP0}
\end{align}
It will be convenient to recast this in terms of the (non-equilibrium) free energy, given by:
\begin{align}
    F(t) = \langle E \rangle_{P(\bm{\sigma},t)} - TS_{\textrm{sys}}(t)
    \label{eq:FE0}
\end{align}
Combining Eqs.~\eqref{eq:EP0} and \eqref{eq:FE0} and integrating over a time window $[t_0,t_f]$ leads to the form of  irreversible entropy production on which we will focus  in this paper:
\begin{align}
    \Delta S_{\textrm{tot}} &= \beta (W_{t_0\rightarrow t_f} - \Delta F) \geq 0.
\end{align}
where $\beta = T^{-1}$. Under quasistatic driving, the total work vanishes, and the entropy produced is just the change in free energy. Below we will be interested in finite time driving strategies which incur additional dissipation. Although we defined the above expressions at the ensemble level, the same relations admit trajectory level descriptions when appropriately defined \cite{Van_den_Broeck_2010,Van_den_Broeck_2015}. 

\section{Equilibrium Behaviour of Memory Networks}
\label{sec:equilibrium}

We are  interested in the dynamics and thermodynamic cost associated with polynomial DenseAMs.  To establish methods, we  first characterize stationary distributions and equilibrium free energies. These results will be useful when we examine network relaxation because  we can calculate final equilibrium free energies exactly in the large system limit. We assume that the network interacts with an infinite bath at inverse temperature $\beta$. With no external fields, the stationary distribution satisfies:
\begin{align}
    P_{\textrm{eq}}(\bm{\sigma}) &= \frac{1}{\mathcal{Z}}\exp[-\beta \mathcal{H}(\bm{\sigma})]= \frac{1}{\mathcal{Z}}\exp\left[\frac{\beta}{N^{k-1}}
    \sum_\mu (\mathbf{\bm{\sigma}}\cdot \bm{\xi}^\mu)^k \right] \, .
    \label{eq:Gibbs}
\end{align}
Define the total \textit{alignment} of the network state with memories $\bm{\xi}^\mu$ as 
\begin{equation}
    \phi^\mu =\frac{1}{N} \bm{\sigma}\cdot \bm{\xi}^\mu = \frac{1}{N}\sum_i (\sigma_i\xi^\mu_i)
    \label{eq:Def1}
\end{equation}
Since $\sigma_i = \pm 1$ and $\xi_i^\mu = \pm 1$, alignment lies between $-1$ and $1$.
Notably, the energetics depend solely on these alignments; however, the entropics in general may not.
We will consider memory loads below saturation $p \ll \alpha^\star N^{k-1}$.  
Importantly, for quadratic networks at equilibrium 
in the absence of external fields, the entropy and free energy can  be expressed in purely in terms of the alignments at large $N$, without any microscopic details of the system. However, near saturation additional spin glass degrees of freedom are necessary  \cite{Sompolinsky2,Gardner}.  We will avoid  saturation here; for details of that regime see  \cite{Gardner, StudentTeacher}. The quadratic (Hopfield) case is described in \cite{Sompolinsky1,Sompolinsky2}. 

The starting point for understanding the equilibrium behaviour of the general model is the partition function:
\begin{align}
    \mathcal{Z} &= \sum_{\{\bm\sigma\}} \exp[\mathcal{H}(\bm\sigma)
    = \sum_{\{\bm\sigma\}} \exp[\frac{\beta}{N^{k-1}}\sum_\mu (\bm\xi^\mu \cdot \bm\sigma)^k] 
\end{align}
For now, we assume that there are no external fields. We  proceed by inserting delta functions enforcing the definition of the alignments: 
\begin{align}
    \mathcal{Z} &= \sum_{\{\bm\sigma\}} \exp[\frac{\beta}{N^{k-1}}\sum_\mu (\bm\xi^\mu \cdot \bm\sigma)^k] 
    = C \, \sum_{\{\bm\sigma\}} \int \prod_\mu d\phi^\mu  \, 
                \delta(\phi^\mu- \frac{1}{N}\bm\xi^\mu \cdot \bm\sigma)
                \exp[N\beta \sum_\mu (\bm\phi^\mu)^k]
    \\ 
    & = C \, \sum_{\{\bm\sigma\}} \int D[\phi^\mu,\tilde{\phi}^\mu] \,  e^{N\sum_\mu 
                \tilde{\phi}^\mu (\phi^\mu-\frac{1}{N}\bm\xi^\mu \cdot \bm\sigma)+N\beta \sum_\mu (\bm\phi^\mu)^k}       
\end{align}
Here $C$ absorbs overall constants  that play no role in the analysis, and $D[]$ is  shorthand for the integral measure. We get an additional set of conjugate fields $\tilde{\phi}^\mu$ via the standard integral representation of the delta functions along contours on the imaginary axis from $-i\infty$ to $+i\infty$.
As in the quadratic case, (\cite{Sompolinsky1,Sompolinsky2}), the spins decouple from each other, and instead collectively couple to the mean fields $\phi^\mu$ and $\tilde{\phi}^\mu$. As such, we can explicitly perform the sum over spin states $\{\bm{\sigma}\}$, and write the partition function in terms of an effective action (see Appendix):
\begin{align}
    \mathcal{Z} &= 
        C \int D[\phi^\mu,\tilde{\phi}^\mu] \,  e^{-N\mathcal{S}[\bm\phi,\bm \tilde{\bm\phi};\{\xi^\mu\}]}
        \label{eq:DANEffectiveZ}
        \\ 
       \mathcal{S} &= - \sum_\mu \tilde{\phi}^\mu\phi^\mu - {1 \over N}
        \sum_i^N \ln\cosh(\sum_\mu \tilde{\phi}^\mu \xi^\mu_i) -
        \beta \sum_\mu (\phi^\mu)^k
        \label{eq:DANEffectiveAction}
\end{align}

\begin{figure}[t]
    \centering
    \includegraphics[width=.95\textwidth]{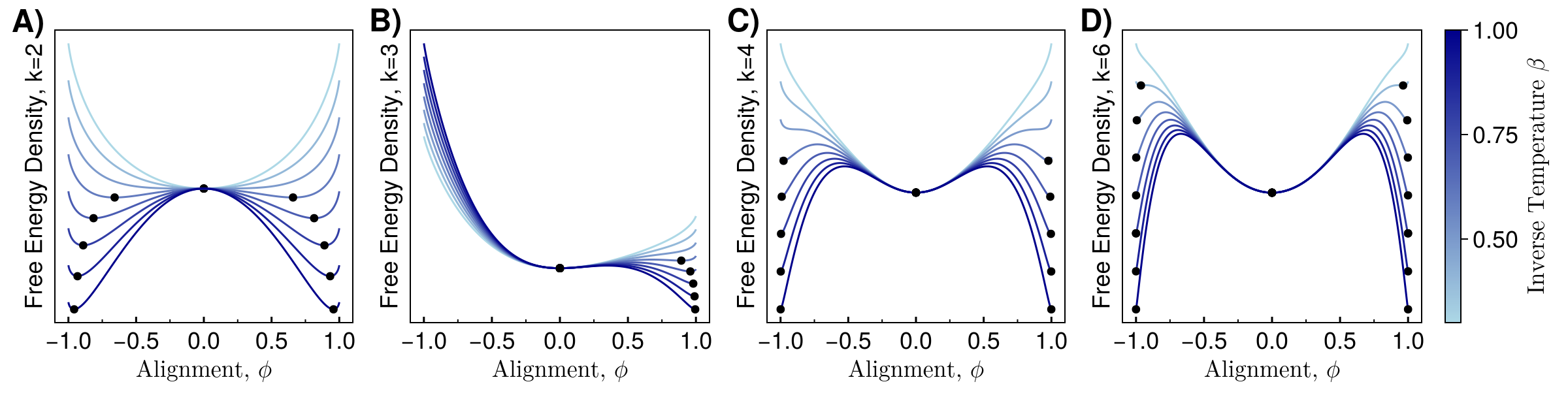}
    \caption{ The free energy landscape of  single memory polynomial DenseAM network as a function of memory alignment for  \textbf{(A)} $k=2$ (Hopfield) \textbf{(B)} $k=3$, \textbf{(C)} $k=4$, \textbf{(D)} $k=6$ networks, and  various temperatures (lighter colors = higher temperature (smaller $\beta$)). For $k=2$, the free energy landscape is identical to that of the mean field Ising model. In this case, at low temperature (large $\beta$) the free energy has aligned and anti-aligned ($\phi=\pm 1$) minima, and an unligned ($\phi=0$) maximum, while at high temperature (small $\beta$) the only minimum is unaligned ($\phi=0$).  
    For $k>2$ there is always a local minimum of the free energy at zero alignment for any finite temperature, leading to a spurious stored memory.  However, the minima associated with true memory alignment are  closer to $\phi = \pm 1$ for the higher order networks at comparable temperature, implying that the memory is  more accurately stored in the free energy basin.  The walls of the energy valley surrounding the stored memory are steeper for larger $k$; so  dynamics that drives an initial state to a free energy minimum will be able to correct a narrower range of errors in alignment of the initial state with the true memory.}
    \label{fig:FreeE}
\end{figure}

At finite $N$, the distribution in $P(\phi^\mu)$ has a width that scales like $\sqrt{N}$. In the  $N\rightarrow{\infty}$ limit the integrand localizes. So the logarithm of the integral in Eq.~\ref{eq:DANEffectiveZ} becomes identically equal to the effective action evaluated at the saddles in the alignments $\bm\phi$ and their conjugates $\bm{\tilde{\phi}}$ (see Appendix). In the thermodynamic limit, the action evaluated at these saddles reproduces the free energy and we find a self consistency equation in the alignments:
\begin{align}
    \phi^{\mu *} 
    &=
    \mathbb{E}_{\bm{x}^\nu} \left[\tanh\Big(k\beta \, \Big[(\phi^{\mu*})^{k-1}+\sum_{\nu\neq\mu} (\phi^{\nu*})^{k-1}x^\nu \Big]\Big) \right] \, ,
    \label{eq:k-consistency2}
\end{align}
where each $x^\nu=\pm 1$ with equal probability. If the network stores $p$ memories, then there are $p$ such alignments. However, only $\mathcal{O}(1)$ can be nonvanishing as $N$ grows large in the thermodynamic limit (see Appendix). As a result, only a small number of degrees of freedom are needed to understand the self consistency equation (Eq.~\ref{eq:k-consistency2}), and by extension, the free energy at equilibrium, so long as the number of memories does not grow too quickly with system size $p<\mathcal{O}(N^{k-1})$. 

Before moving to the dynamics of the system, consider the network storing a single memory, as it will hint towards a failure mode of these networks that is observed dynamically. In the thermodynamic limit, the free energy can be understood purely in terms of the alignment with a single memory (see Appendix):
\begin{align}
    \beta f(\phi) = \mathcal{S}[\phi,\tilde{\phi}]|_{\tilde{\phi}^*}
    &= -\beta\phi^k + \frac{1}{2}[(1-\phi)\ln(1-\phi)+(1+\phi)\ln(1+\phi)]
    \label{eq:DANfreeenergy}\\
    \mathcal{Z} &= C \int d\phi e^{-N f(\phi)}
\end{align}
For the quadratic Hopfield models ($k=2$) with a single memory, the free energy in (\ref{eq:DANfreeenergy})  is identical to the mean field Ising model free energy density.

We can find the equilibrium configurations that dominate the partition function by minimizing the free energy.  At low temperature (large $\beta$), a short calculation shows the quadratic Hopfield model ($k=2$)  has two minima corresponding to the aligned and antialigned states, as it is identical to the mean field Ising model. We are interested in  general polynomial DenseAM networks with $k>2$. As seen in Fig.~\ref{fig:FreeE}, the free energy for these models has global minima at aligned (and antialigned if $k$ is even) states with $\phi \sim \pm 1$, as well as a local minimum at $0$ that is present at any finite temperature. To understand the origin of this local free energy minimum, we can Taylor expand the free energy $f$ near zero alignment:
\begin{align}
    f(\phi) &= -\phi^k + \frac{1}{2\beta}[(1-\phi)\ln(1-\phi)+(1+\phi)\ln(1+\phi)]
    \\ 
    &\sim 
    -\phi^k + \frac{1}{\beta}[\phi^2/2 + \phi^4/12 + ...]
\end{align}
For $k=2$, this becomes concave down at zero when $\beta>\frac{1}{2}$, and the extremum at zero becomes unstable.  However, for $k>2$, the local free energy extremum at zero alignment is always concave up, and thus represents a local free energy minimum (Fig.~\ref{fig:FreeE}).  This local minimum plays a crucial role in the dynamics of these networks at finite temperature as we will discuss in the next section;  it will act as an attractor under dynamics that satisfy detailed balance at equilibrium.  

Physically, the energy landscapes become flatter near zero alignment, but steeper near alignment with a memory, as $k$ increases.
As such, at larger $k$, thermal fluctuations can preferentially walk the state of the system along
the flat regions of the energy landscape at finite temperature when the initial state is not well-aligned with a memory.  In the next section we will show that this is a source of a finite temperature dynamical instability that has not been studied before. Conversely, when the system is aligned with a memory, the system will fluctuate less at finite temperature because the energy barriers are steeper and so the memories are more stable.   Put differently, lower order networks have larger basins of attraction for stored memories and hence can perform reconstruction of initial states that are  more corrupted ($\phi$ starts closer to zero), but in exchange will have have larger fluctuations/errors in the reconstructions themselves.  We will show this  explicitly when we turn to the dynamics of the network.

\section{Dynamics and Stochastic Thermodynamics}
\label{sec:dynamics}

Having understood the system at equilibrium, we want to explore different modes of network operation, their dynamics, and the resulting thermodynamic costs. Suppose
we have a set of $p$ memories $\{\bm{\xi}^\mu\}_{\mu=1}^p$ stored by a network. Additionally, we have a set of $q$ corrupted patterns, $\left(\bm{\zeta}^1,...,\bm{\zeta}^q\right)$ corresponding to some pattern stored by the network; each $\bm{\zeta}$ is obtained by flipping $\gamma N$ spins in some memory $\bm{\xi}$,  .  We will call $\gamma$ the {\it corruption fraction}. As we will demonstrate, the alignments are sufficient to fully characterize both the dynamics and the thermodynamics in the large $N$ limit for \textit{arbitrary} driving strategies built from partial memories. 

\subsection{Relaxation Dynamics}
\label{sec:relaxation}
First consider the simple case where the network starts perfectly localized to a single partial memory $\bm{\zeta}$, and is allowed to relax spontaneously; i.e., we perform no work. This is the standard mode in which these networks are used, though ordinarily they are understood at zero temperature under greedy descent dynamics. At finite temperature, we will see that there is a failure mode of the higher order networks that was not previously described. We will also characterize associated tradeoffs between reconstructability and reconstruction loss. If we want to pattern complete multiple corrupted patterns, there is also additional thermodynamic cost incurred in the form of work, to which we will return to in a later section.

We start by demonstrating that the dynamics in the alignments close. To this end, we start with the dynamics of the expected spin state, which follows from the master equations:
\begin{align}
    \partial_t\langle \sigma_i \rangle &= \partial_t \sum_{\bm{\sigma}} P(\bm{\sigma},t)\sigma_i =\sum_{\bm{\sigma}} \sum_j \left[ \sigma_i \Gamma_j(S_j\bm{\sigma};t)P(S_j\bm{\sigma};t)
    -\sigma_i\Gamma_j(\bm{\sigma};t)P(\bm{\sigma};t) \right]\\
    &=- \frac{1}{\tau} \langle \sigma_i\rangle
    + \frac{1}{\tau} \langle \tanh(\frac{1}{2}\beta
    \sigma_i\Delta_i\mathcal{H})\rangle
    \label{eq:masterAvSpin}
\end{align}
where $S_j$ is an operator flipping spin $j$ and $\Gamma_j$ describes transition rates between states with $\sigma_j$ flipped, and $\Delta_i \mathcal{H}$ is the change in the Hamiltonian from flipping spin $i$.  The first line follows simply by taking the expectation value of $\sigma_i$ in the master equations (\ref{eq:Master1}).  To get the second line we use the transition rates specified in (\ref{eq:Master2}) and carry out the spin sums using three facts: (a) the spins take values $\pm 1$, (b) $\tanh$ is an odd function of its argument, and (c) $\Delta_j \mathcal{H}$ changes sign when evaluated on a configuration with flipped $\sigma_j$.  In what follows, we will set $\tau = 1$.
The change in the Hamiltonian from a single spin flip is given by:
\begin{align}
\Delta_i\mathcal{H} &= 
- \frac{1}{N^{k-1}}\sum_\mu [(S_i(\bm\sigma \cdot \bm\xi^\mu))^k-
    (\bm\sigma \cdot \bm\xi^\mu)^k] + 2 h_i \sigma_i\\
&= - \frac{1}{N^{k-1}}\sum_\mu [(N\phi^\mu-2\sigma_i\xi_i^\mu)^k-
    (N\phi^\mu)^k] + 2 h_i \sigma_i \\
&\approx 2\sum_\mu [k(\phi^\mu)^{k-1}\sigma_i\xi_i^\mu] + 2 h_i \sigma_i
\label{eq:EnergyDifference}
\end{align}
The first equality arises by explicitly evaluating the change in the DenseAM network Hamiltonian (\ref{eq:Hamil2}) when one spin is flipped, the second equality applies the definition from above that $(1/N) \bm{\sigma} \cdots \xi^\mu = \phi^\mu$, and the third line keeps the leading terms in powers of $N$ which are the relevant ones for the large $N$ limit of interest to us.  So, after inserting the change in the Hamiltonian (\ref{eq:EnergyDifference}) into the evolution equation for individual spin expectation values (\ref{eq:masterAvSpin}) we find that:
\begin{align}
    \partial_t\langle \sigma_i\rangle 
    &= -\langle\sigma_i\rangle
    +\Big\langle\tanh\Big[ k\beta\sum_\nu  (\phi^\nu)^{k-1}\xi_i^\nu + \beta h_i\Big]\Big\rangle
    \label{eq:meanSpin}
\end{align}
where $\sum_i\sigma_i\xi_i^\mu = N\phi^\mu$. 
Note that the nonlinear second term in (\ref{eq:meanSpin}) couples the dynamics of individual spin expectation values to  spin correlations of all orders.  Thus, to completely determine the dynamics we have to solve simultaneously for all $2^N$ possible correlation functions of the $N$ spins. 
We can similarly write dynamical equations for the correlation of the spins in any given subset $\mathcal{A}$:
\begin{equation}
\partial_t\langle \prod_{i\in \mathcal{A}}\sigma_i\rangle 
=
- \langle \prod_{i\in \mathcal{A}}\sigma_i\rangle
    + \langle\sum_{j\in \mathcal{A}}\tanh(\frac{1}{2}\beta\prod_{i\in \mathcal{A}}\sigma_i\Delta_j\mathcal{H})\rangle
=
- \langle \prod_{i\in \mathcal{A}}\sigma_i\rangle
    +\langle\sum_{j\in \mathcal{A}}\prod_{i\in \mathcal{A},i\neq j}\sigma_i\tanh(\frac{1}{2}\beta
    \sigma_j\Delta_j\mathcal{H})\rangle
    \label{eq:sigmaevol}
\end{equation}

\begin{figure}[t]
    \centering
    \includegraphics[width=.95\textwidth]{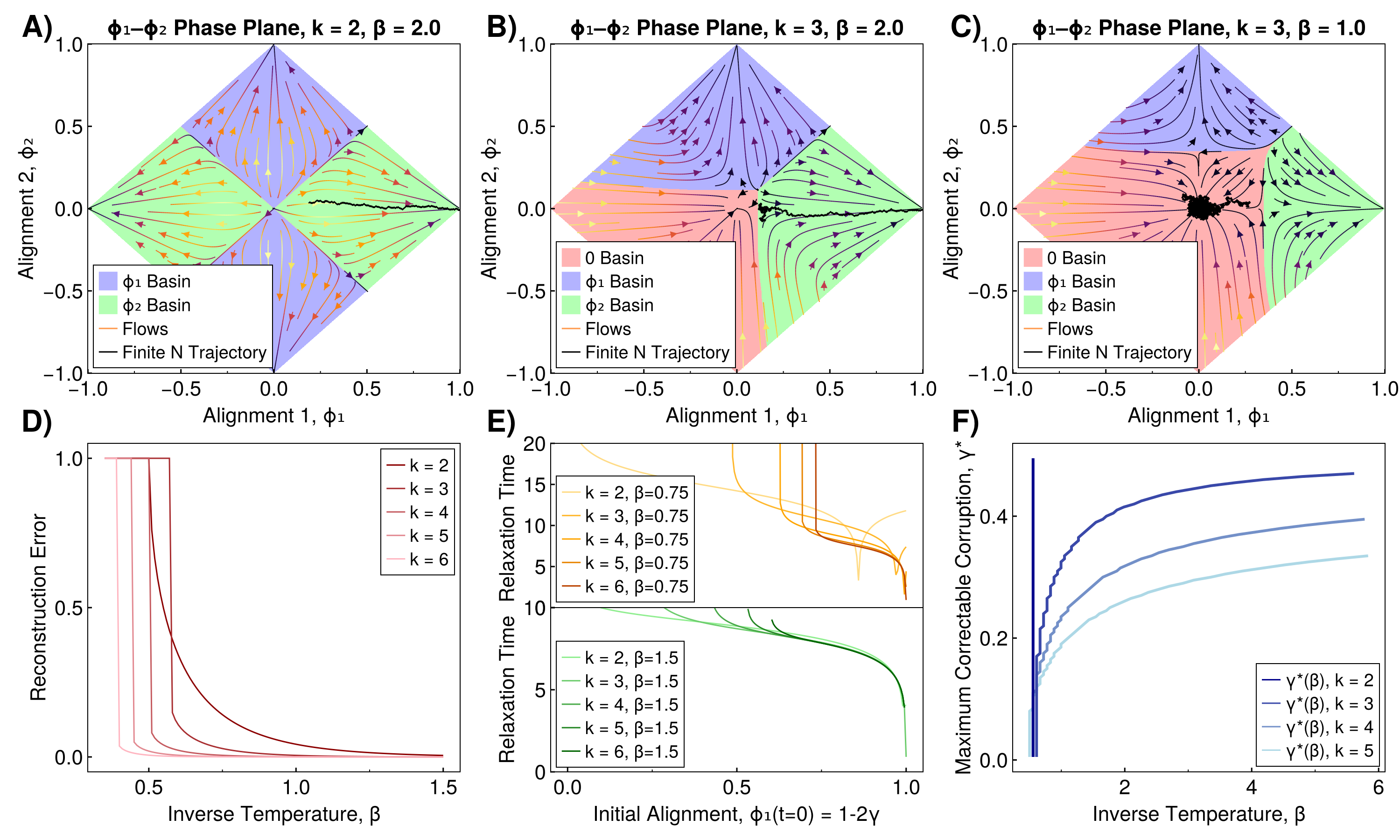}
    \caption{
    \textbf{(A-C)} Phase Portraits associated with two alignments for DenseAM networks  storing two memories, with relaxation dynamics given by Eq.~\ref{eq:RelaxationDynamics}. \textbf{{(A)}} The quadratic (Hopfield) network at low temperature ($\beta=2.0$). \textbf{{(B-C)}} The cubic network at \textbf{(B)} low temperature ($\beta=2.0$) and \textbf{(C)} intermediate temperature ($\beta=1.0$). Given an initial state $\vec{\phi}(t=0)$, colors indicate which which attractor the dynamics drive the state towards.
    These correspond to partial alignment (or anti alignment for $k$ even) with each memory, and zero alignment for $k>2$.  For $p>2$, additional attractors associated with linear combinations of memories also appear. In black are single trajectories associated with finite $N$ Glauber simulations.
    Leading order corrections to deterministic flows are derived in the appendix.
    \textbf{(D)} The reconstruction error $1-\phi_{eq}$ after relaxation as a function of $\beta$ for DenseAM networks with varying nonlinearities, assuming relaxation is successful. Higher order networks reconstruct memories with greater fidelity when reconstruction is successful.
    \textbf{(E)} Time taken to relax to within $\epsilon=10^{-4}$ of dynamic fixed points for DenseAM networks with varying nonlinearities, as a function of initial state corruption and at two different temperatures. This relaxation time grows approximately logarthmically in $\gamma$ and in $\epsilon$.  \textbf{(Top)} At intermediate temperatures $\beta=.75$, higher order networks relax more quickly in the regime where relaxation is successful. As temperature decreases \textbf{(Bottom)}, relaxation times become similar, as the tanh term in Eq.~\ref{eq:RelaxationDynamics} approaches a step function.
    \textbf{(F)} Plot of the maximum amount of corruption that DenseAMs can correct at various temperatures. Lower order networks can correct patterns that are more highly corrupted, although at lower fidelity as in \textbf{(D)}. 
    }
    \label{fig:dynamics}
\end{figure}

Here, we are interested in the dynamics of the alignments $\phi^\mu = (1/N) \bm{\sigma} \cdot \bm{\xi}^\mu$. 
By multiplying (\ref{eq:sigmaevol}) by $\xi^\mu_i$  and summing over $i$ we find  that
\begin{align}
    \partial_t \langle  \phi^\mu \rangle &= -\langle \phi^\mu \rangle
    + \frac{1}{N}\sum_i \xi_i^\mu\Big\langle \tanh\Big[ k\beta\sum_\nu  (\phi^\nu)^{k-1}\xi_i^\nu + \beta h_i\Big]\Big\rangle
    \\
    &= -\langle \phi^\mu \rangle + 
    \frac{1}{N}\sum_i \Big\langle \tanh\Big[ k\beta(\phi^\mu)^{k-1} + k\beta\sum_{\nu\neq\mu}  (\phi^\nu)^{k-1}\xi_i^\mu\xi_i^\nu + \beta\xi_i^\mu h_i\Big]\Big\rangle
    \label{eq:PhiDin}  
\end{align}
As with single spins, the $\tanh$ nonlinearity in  (\ref{eq:PhiDin}) couples the dynamics of the expectation value of $\phi^\mu$ to higher order correlation functions in the alignments. However, at large $N$ and low temperature the probability distribution in $\bm\phi$ will be localized to its mean, with an expected width in the distribution at each time that is $\mathcal{O}(\frac{1}{\sqrt{N}})$ due to the law of large numbers. As in the equilibrium case, the stochastic contribution from non-aligned patterns is expected to vanish so long as $p<<\mathcal{O}(N^{k-1})$. 
Moreover, in this low load regime the equilibrium alignments are not stochastic in the large $N$ limit, as the contribution of unaligned patterns is vanishing. This implies that the dynamics of any fluctuations in the alignments must be overdamped, and relax.
While this reasoning is heuristic, the argument is made exact in the Appendix via the Kramers-Moyal expansion  for the more restricted case when $p<\mathcal{O}(\sqrt{N})$. In any case, the dynamics in the alignments become deterministic, and are given, without the presence of external fields, as:
\begin{align}
    \partial_t  \phi^\mu  &=  -\phi^\mu  + 
    \mathbb{E}_{\bm{x}}\tanh\Big[ k\beta(\phi^\mu)^{k-1} + k\beta\sum_{\nu\neq\mu}  (\phi^\nu)^{k-1}x^\nu\Big]
    \label{eq:RelaxationDynamics}  
\end{align}
where $x =\pm 1$ with equal probability.  Here we noted that $\xi_i^\mu \xi_i^\nu = \pm 1$ with equal probability because the memories are uncorrelated, and then used the central limit theorem to replace the sum on $i=1 \cdots N$ with an expectation value on $x$ for large $N$.

Now as in the equilibrium case, only $\mathcal{O}(1)$ alignments can have $\mathcal{O}(1)$ magnitude at any given time, 
with the remaining vanishing in the thermodynamic limit. The general reason for this is that we have assumed that the stored memories are random vectors in the high dimensional space of spin polarizations, and their number is sub-exponential in $N$. Then at large $N$, $\vec{\xi}^\mu\cdot\vec{\xi}^\nu\simeq \mathcal{O}(\sqrt{N})$ for $\mu\neq \nu$ because random $N$ dimensional binary vectors have vanishing overlaps distributed with zero mean and standard deviation $\sqrt{N}$.   

To understand why, suppose that $\vec{v}$ is some binary vector.  If $\vec{v}$ is fully aligned with $\vec{\xi}^1$, then $\vec{v}\cdot\vec{\xi}_1=N$ and its dot product with the other patterns will be $\mathcal{O}(\sqrt{N})$. If we quantify alignment with pattern $\mu$ as  $\frac{1}{N}\vec{v}\cdot \xi^\mu$, then the criterion for nonvanishing alignment is that $\vec{v}\cdot\xi^\mu$ has a term scaling at least as fast as $N$.  For example, suppose that for half the indices $\vec{v}$ is aligned with pattern 1, and the other half of its indices it is aligned with pattern 2. Then by similar reasoning as above, $\vec{v}\cdot \vec{\xi}^1 \simeq N/2 \pm \sqrt{N/2}$ and $\vec{v}\cdot \vec{\xi}^2 \simeq N/2 \pm \sqrt{N/2}$, and the remaining alignments will again all be $\mathcal{O}(\sqrt{N})$.  Likewise, suppose $\vec{v}$ is perfectly aligned with each of a set $k$ memories $\vec{\xi}^i$ in a fraction $\mathcal{O}(1/k)$ of the spins, then $\vec{v}\cdot\xi_i \simeq N/k \pm\sqrt{N/k}$ for $i\in\{1,...,k\}$, and the remaining  alignments must be $\mathcal{O}(\sqrt{N})$.  In general, for the spin state $\vec{v}$ to have $\mathcal{O}(1)$ alignment with some memory we must have $\mathcal{O}(N)$ aligned spins.  So, since there are only $N$ spins in total, we can at most align the spin state with $\mathcal{O}(1)$ different memories.

As in the equilibrium case, the sum over the nonaligned patterns only contributes if the number of stored patterns grows as fast as $p\sim \mathcal{O}(N^{k-1})$.
Since we consider memory loads below this bound,  we can discard the sum over nonaligned patterns for understanding the dynamics of $\phi^\mu$ with $\mu\in S$. This leads to a set of finitely many differential equations.
As such, it suffices to understand the dynamics of networks storing $\mathcal{O}(1)$ memories to understand the dynamics of networks storing $p<\mathcal{O}(N^{k-1})$ memories.
When considering loads near the capacity of the network, the sum over nonaligned patterns  becomes nonvanishing, and adds a stochastic component to the dynamics.
We do not consider this here, but this stochastic contribution can potentially be approximated by analyzing the complete generating functional for the dynamics, which encodes full path probabilities of the system and allows systematic calculation of dynamical correlationa. We can also subsequently include dTAP-like reaction terms \cite{dTap}which provide corrections to mean-field dynamics by capturing feedback from fluctuations. We leave such extensions for future work.

We can numerically integrate Eq.~\ref{eq:RelaxationDynamics} using a small number of degrees of freedom to understand the behaviour of these networks. \textbf{This leads to our first set of results.} As expected from the equilibrium free energies, the dynamics exhibit an attractor at zero alignment when $k>2$, which is shown for the two memory case in Fig.~\ref{fig:dynamics}. The size of this spurious state depends strongly on temperature, and vanishes as $\beta\rightarrow \infty$. However, this leads to a potential failure mode for the higher order networks that is not observed in the quadratic network, where pattern completion cannot be achieved. There are additional spurious attractors at finite temperature associated with linear combinations of multiple memories when the number of memories is $p\geq 3$,
consistent with results known for the Hopfield network \cite{Sompolinsky1, Sompolinsky2}. 

When starting with an inital state that is a random corruption of a memory, the probability that $\bm{\phi}$ starts in an attractor associated with these additional spurious states is vanishing as $N\rightarrow \infty$. This is because these spurious states require simultaneous, non-vanishing alignment with multiple memories. Even under strong corruption, if the corruption is unstructured, the initial state retains an $\mathcal{O}(1)$ alignment with the memory being corrupted away from, while overlaps with all other memories are only order $\mathcal{O}(\frac{1}{\sqrt{N}})$. 
Although the higher order networks have an additional attractor at zero alignment, when they do relax to the correct memory, they typically relax faster and reconstruct memories with fewer errors (Fig.~\ref{fig:dynamics}), as suggested by the qualitative analysis of the free energy in the previous section.
These relaxation  patterns suggest that for a memory to be reconstructed with given error rate / corruption fraction, the higher order networks must be operated at lower temperatures so that they do not relax towards the spurious state at zero alignment. As we will see below, this will lead to higher energy dissipation, as the entropy produced is inversely proportional to temperature (Eq. ~\ref{eq:EPDensity}).

\subsubsection{Relaxation Thermodynamics}
\label{sec:relaxthermo}
As previously mentioned, over a finite time interval $[t_0,t_f]$, the (irreversible) entropy produced is given by the work performed by the network and the change in free energy over that time interval:
\begin{align}
    \Delta S_{tot} &= \beta (W_{t_0\rightarrow t_f} - \Delta F)
\end{align}
We are interested in entropy and work densities, $\Delta s_{tot}=\frac{1}{N}\Delta S_{tot}$ and $w=\frac{1}{N}W$ as we take $N\rightarrow\infty$. For the simple relaxation described in this subsection, the work is zero, and the only relevant thermodynamics are due to changes in free energy.

\begin{figure}[t]
    \centering
    \includegraphics[width=.99\textwidth]{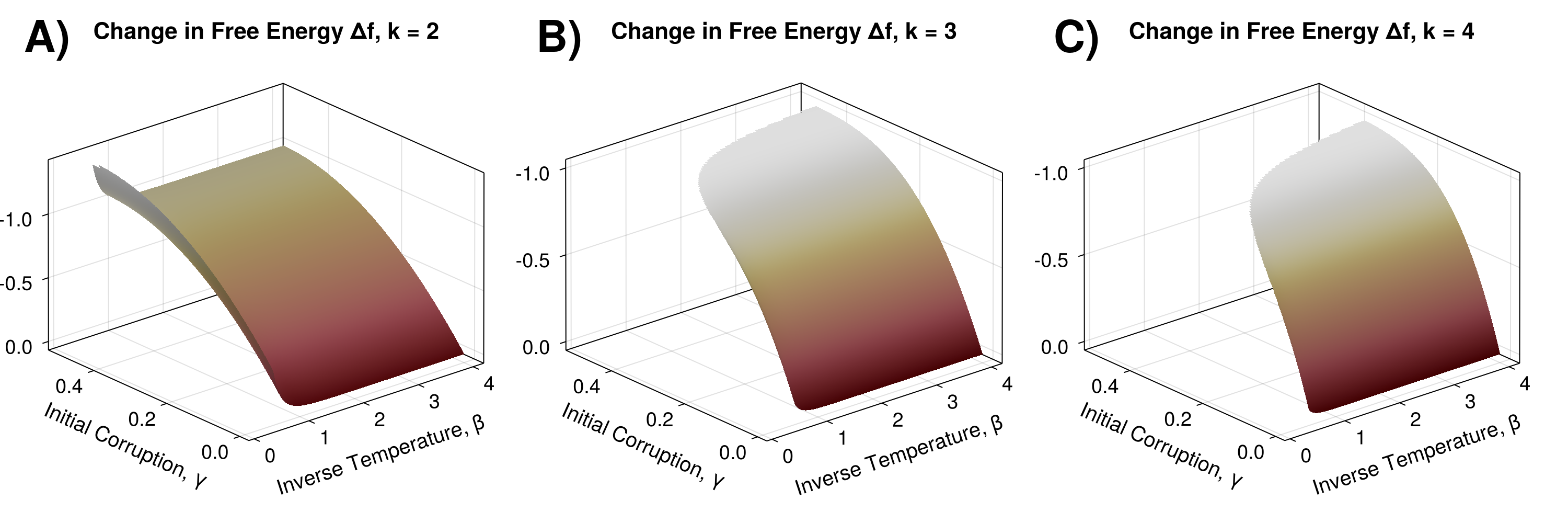}
    \caption{
    The change in free energy density in the \textbf{(A)} $k = 2$, \textbf{(B)} $k = 3$, and \textbf{(C)} $k = 4$ DenseAM networks  as they relax from a corrupted pattern to the equilibrium distribution around the reconstructed pattern, when such reconstruction is successful, as a function of inverse temperature $\beta$ and initial corruption $\gamma$. Multiplying by $\beta$ reproduces Eq.~\ref{eq:EPDensity}.
    }
    \label{fig:Pringles}
\end{figure}

Before relaxation, the system is localized at the configuration $\bm{\zeta}^1$, i.e., $p_0$ is a delta function centered at that configuration.
So the initial free energy simply equals the energy, given by the Hamiltonian evaluated at $\bm\sigma = \bm{\zeta}^1$. At sufficiently low temperature, and assuming the corrupted pattern does not start too far away from the true pattern (i.e., relaxation succeeds), the final free energy is given by the equilibrium free energy evaluated at the equilibrium allignment, as calculated in the previous section (Fig.~\ref{fig:Pringles}):
\begin{equation}
  \Delta s_{tot} = - \frac{1}{N}\beta\Delta F = -\frac{1}{N}\beta [F(t_f)  - F(t_0) ]
    = - [\mathcal{S}[\bm{\phi}^\star,\bm{\tilde{\phi}}^\star;\bm\xi,\beta]  
    -\frac{\beta}{N^k} \sum_\mu(\bm{\zeta}^1\cdot \bm{\xi}^\mu)^k ]-\ln(2) \, \label{eq:EPDensity}
\end{equation}
The change in free energy is just the action evaluated at the final state, minus the Hamiltonion evaluated at the initial state, along with an additive factor $\ln 2$. This term originates from a constant contribution to the equilibrium free energy (i.e. $Z_{eq} = 2^N \times\textrm{[Interacting Part]}$) which is typically dropped, as it plays no role in comparisons between equilibrium free energies. However, it is restored here to ensure consistency with the definition of nonequilibrium free energy, where absolute entropy, and hence additive constants in the free energy at equilibrium, affect the comparison. This constant $\ln 2$ is combinatorially fixed by the full count of equilibrium microstates.

\subsection{Dynamics and Thermodynamics of Driven Networks}
\label{sec:driven}
In the previous section, we characterized how DenseAM networks relax at finite temperature. During such relaxation, no work is done on the system, and the only entropy produced is associated with heat dissipated to the bath. We will now consider the DenseAM networks driven over finite durations by external fields $\bm{h}(t)$. In particular, we are interested in the work required to present the network with corrupted memories that are then dynamically corrected.

Different choices for $\bm{h}(t)$ can be viewed as different control strategies, and  we want to choose a protocol $\bm{h}$ which quickly and accurately reproduces each memory from a corrupted sequence $\{\bm{\zeta}^1,...,\bm{\zeta}^q\}$.  We constrain $\bm{h}(t)\in\textrm{Span}\{\bm{\zeta}^1,...,\bm{\zeta}^q\}$, as we assume that an operator using the network only has knowledge of the partial memories.  We assume that each $\bm{\zeta}$ corresponds to a memory stored by the network, with a fraction $\gamma$ of the spins flipped. So  we write $\zeta_i^\mu = C_i^\mu \xi_i^{\mu}$, where the $C_i^\mu$ are independent random variables which take the values $-1$ and $1$ with probability $\gamma$ and $1-\gamma$ respectively.  For simplicity, we assume that there is no more than one partial pattern $\bm{\zeta}$ associated with each true pattern, but generalizing to multiple partial patterns associated with single memories is straightforward. With these assumptions the driving fields $\bm{h}(t)$ can  be expressed in terms of control variables $u(t)$ as: 
\begin{align}
    h_i(t) &=\sum_\mu u^\mu(t)\zeta^\mu_i =\sum_\mu u^\mu(t) C_i^\mu \xi_i^{\mu}    
    \label{eq:controlvars}
\end{align}
We can include this external field in the dynamical equation for the mean alignments (\ref{eq:PhiDin}).  

\begin{figure}[t]
    \centering
    \includegraphics[width=.99\textwidth]{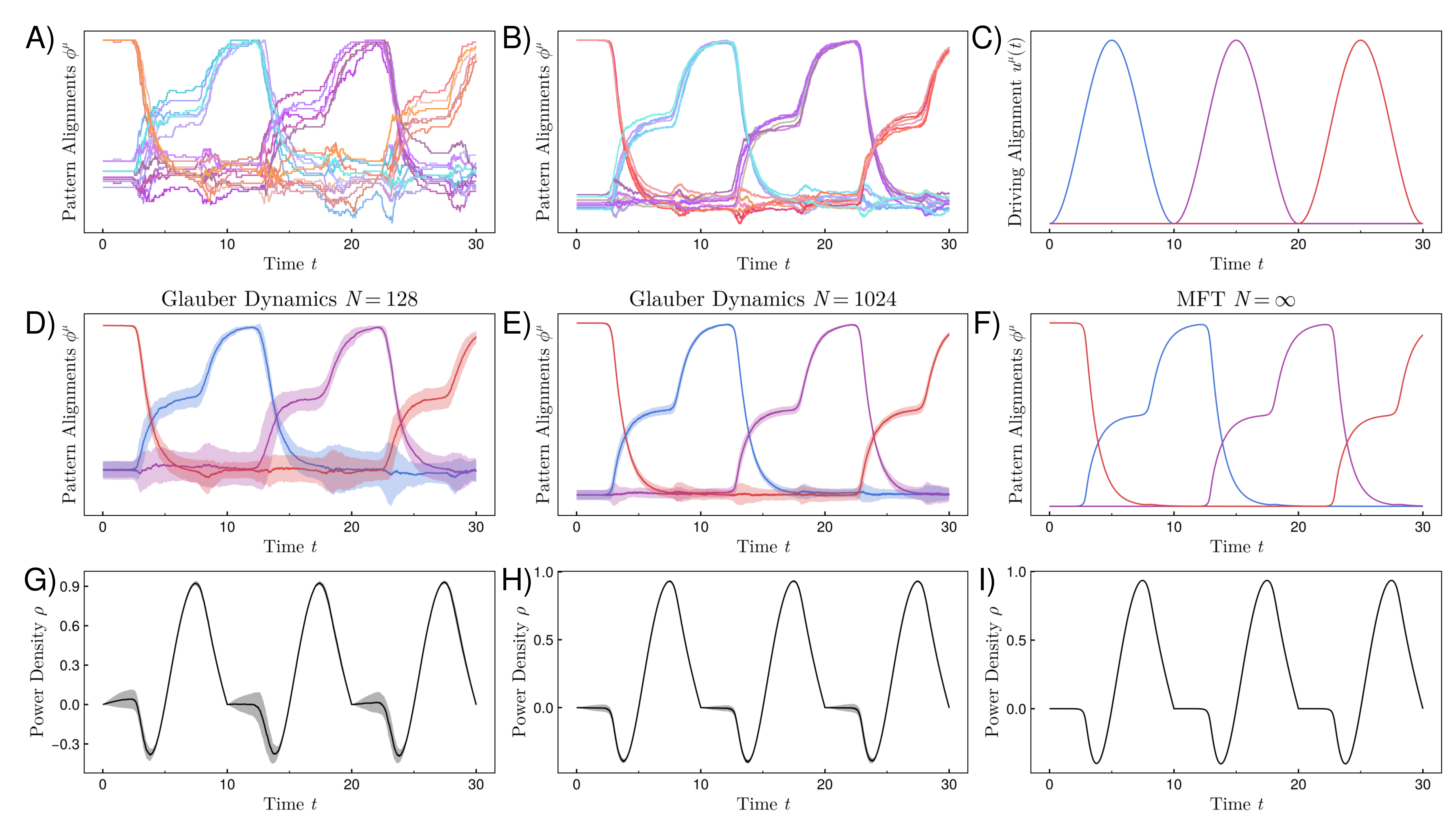}
    \caption{ 
    Numerical Demonstration of mean field theory for $k=3$ networks with 3 memories. Glauber simulations for \textbf{(A)} $N=128$ and \textbf{(B)} $N=1024$ Neurons under corrupted driving strategy \textbf{(C)} ($\gamma = .25$) (plotted are alignments with each of the three memories). The mean and variance of trajectories for each are shown in \textbf{(D)} and \textbf{(E)}, with the mean field trajectory shown in \textbf{(F)}. As $N$ increases, we expect variances in these trajectories to shrink like $1/\sqrt{N}$. The power density consumed and its variances for each of the three cases are shown in \textbf{(G-I)}. Integrating this gives the work divided by $N$. Over any closed cycle, the integral of this quantity must be positive. The mean field work density calculated from Eq.~\ref{eq:PowerDensity22} \textbf{(I)} agrees with that found from simulation at finite $N$ \textbf{(G,H)}.
    }
    \label{fig:MFT}
\end{figure}

Then, by the same arguments as discussed above for relaxation without driving fields, at large $N$, low temperature, and if the number of stored memories is below capacity, we expect the probability distribution over over alignments to be strongly localized, so that fluctuations around the expectation value will be small.  We can then make a dynamic mean field approximation, and remove the expectation values in (\ref{eq:PhiDin}), treating this expression as a deterministic equation for the mean alignments. The justification for this is identical to that given in the spontaneous relaxation case. We then have:
\begin{equation}
        \partial_t \phi^\mu  = - \phi^\mu + 
    \frac{1}{N}\sum_i \tanh\Big[
    k\beta(\phi^\mu)^{k-1} +
    k\beta\sum_{\nu\neq\mu}  (\phi^\nu)^{k-1}\xi_i^\mu\xi_i^\nu +
    \beta C_i^\mu u^\mu + \beta \sum_{\nu\neq\mu} C^\nu_i \xi_i^\mu\xi_i^\nu u^\nu 
    \Big]
\end{equation}
Finally, recalling  that the memories are assumed to be  uncorrelated we can use the law of large numbers to replace the sum over spins $(1/N) \sum_{i=1}^N$ by expectation values over auxiliary random variables:
\begin{equation}
 \partial_t \phi^\mu  = - \phi^\mu +
  \mathbb{E}_{\bm{Y},\bm{x}}\tanh\Big[
    k\beta(\phi^\mu)^{k-1} +\beta Y^\mu u^\mu +
    \sum_{\nu\neq\mu}  \beta x^\nu[k(\phi^\nu)^{k-1} +
    Y^\nu  u^\nu] 
    \Big]
    \label{eq:PhiDyn3}
\end{equation}
where $x^\mu = \pm 1$ with equal probability and $Y^\mu = -1$ and $+1$ with probabilities $\gamma$ and $1-\gamma$, respectively. We then further partition \eqref{eq:PhiDyn3} into the $\mathcal{O}(1)$ alignments that are nonzero somewhere during a finite time trajectory $[t_0,t_f]$ and those that are unaligned. In the low load regime, only the aligned $\phi^\mu$ contribute to the total dynamics, with stochastic corrections from unaligned patterns vanishing as $N\rightarrow \infty$, as in the undriven case from the last section. This leads to a small set of coupled ODEs which are much simpler to analyze and simulate than repeated Monte Carlo simulations of the complete master equation dynamics at large $N$. Comparisons between these dynamics and finite N CTMC dynamics are shown for a particular driving strategy in Fig.~\ref{fig:MFT}.

 We can now write down an expression for the work done by a particular control strategy $\bm{u}(t)$.  This work is defined in terms of changes in the systems energy levels, weighted by expected occupancy:
\begin{equation}
 \mathcal{W}_{t_0\rightarrow t_f} 
 =
 \int_{t_0}^{t_f}dt \, \langle d_t\mathcal{H}(\bm{\sigma},t)\rangle_{P(\bm{\sigma},t)} 
 =
 \int_{t_0}^{t_f} dt \, 
        \sum_\mu\frac{\partial u^\mu}{\partial t}
        \sum_iC_i^\mu\xi_i^{\mu} \, 
        \langle \sigma_i(t)\rangle 
        \label{eq:work}
\end{equation}
where we arrived at the last expression by using the DenseAM Hamiltonian (\ref{eq:Hamil2}) and the expression for the driving fields in terms of the control variables. 
We assume that the system is initially localized to a single memory and that the control satisfies the boundary conditions $\bm{u}(t_0)=\bm{u}(t_f)=0$. If the system has been successfully driven through a sequence of memories, and is well localized to a single memory at $t_f$, then the change in free energy over the whole trajectory is subextensive in $N$. This is because the leading-order contributions to the free energy at the initial and final states are identical, so only subleading terms, which vanish as $N\rightarrow \infty$, remain. As such, the entropy produced over the whole trajectory is simply the work done multiplied by a factor of $\beta$ under successful driving protocols.  

Next, we integrate the dynamical equation (\ref{eq:meanSpin}) to express the expectation value of the spins in terms of the alignments $\bm{\phi}$:
\begin{align}
    \langle \sigma_i(t)\rangle &= \int_{t_0}^{t} ds \, e^{s-t}
    \langle \tanh\Big[ k\beta\sum_\mu  (\phi^\mu)^{k-1}\xi_i^\mu + \beta h_i\Big] \rangle + e^{-t}\langle\sigma_i(t_0)\rangle
    \label{eq:IntDyn}  
\end{align}
At loads below capacity and as $N\rightarrow\infty$, each alignment $\phi^\mu$ either vanishes, or becomes completely localized around its peak value at each time, obeying the deterministic dynamics of Eq.~\ref{eq:PhiDyn3}, as explained heuristically in the undriven case. As such, we can remove the expectation with respect to the $\tanh$. Additionally, we assume that we have waited a sufficiently long time for the system to equilibrate before doing any work on it, and drop the contribution from the initial state of the system $\bm{\sigma}(t_0)$.  
Inserting the solution for the mean spins back into (\ref{eq:work}), we find:
\begin{equation}
\mathcal{W}_{t_0\rightarrow t_f}
= N\int_{t_0}^{t_f}\rho(t) \, dt ~~~~;~~~~
\rho(t) =
        \frac{1}{N}\sum_\mu \frac{\partial u^\mu(t)}{\partial t}
        \sum_iC_i^\mu\xi_i^{\mu}
        \int_{t_0}^{t} ds \, e^{s-t}
        \tanh\Big[ k\beta\sum_\nu  (\phi^\nu)^{k-1}\xi_i^\nu + \beta h_i\Big] \, .
\end{equation}
Since the $\tanh$ is an odd function of its argument we can pull the factor of $C_i^\mu \xi_i^\mu = \pm 1$ into it. Now using the definition of the driving fields in terms of the control variables (\ref{eq:controlvars}), we can perform similar manipulations as in previous sections:  (a) First we separate the $\nu = \mu$ and $\nu \neq \mu$ parts of the sums inside the $\tanh$;  (b) Second we recognize that, since $\xi_i^\mu$ and $\xi_i^\nu$ are uncorrelated, the law of large numbers says that the effect of the sum on spins $(1/N) \sum_i$ is to replace any occurrence of $\xi_i^\mu \xi_i^\nu$ for fixed $\mu$ with a random variable $x^\nu$ taking values $\pm 1$ with equal probability; (c) Third,   the sum on spins $(1/N) \sum_i$ similarly allows us to replace occurrences of $C_i^\mu$ by a random variable $Y^\mu$ which equals $-1$ with probability $\gamma$ and $1$ with probability $1-\gamma$.  This gives:
\begin{equation}
\rho(t) 
=
    \sum_\mu {\partial u^\mu(t) \over \partial t} 
    \int_{t_0}^{t} ds \, e^{s-t}   
    \mathbb{E}_{\{\bm{Y},\bm{x}\}}
    \tanh\Big[ \beta Y^\mu
    \big[ k(\phi^\mu)^{k-1} +
    \sum_{\nu\neq \mu} x^\nu (k(\phi^\nu)^{k-1}
    +  Y^\nu u^\nu)
    \big]  
    + \beta u^\mu(t)
    \Big]
       \label{eq:PowerDensity22} 
\end{equation}
Along with Eq.~\ref{eq:PhiDyn3}, \textbf {this expression for the instantaneous power (and by extension, work) is our most important result}.

Eq.~\ref{eq:PowerDensity22} demonstrates that finite time thermodynamic quantities can be understood exactly in this system purely in terms of macroscopic states of the system $\bm{\phi}(t)$ and $\bm{u}(t)$.  The expression is exact in the mean field theory limit which applies at large $N$, when the number of memories is sufficiently less than the capacity of the network. As before, only the $\mathcal{O}(1)$ patterns that align somewhere in the finite time trajectory need to be tracked, with the stochastic nonaligned contribution vanishing at low to intermediate load. Finite system size corrections of order $\frac{1}{\sqrt{N}}$ bound the error of this instantaneous power, as demonstrated in Fig.~\ref{fig:MFT}. Interestingly, after eliminating the internal degrees of freedom, we are left with an instantaneous power which depends on the {\it history} of the macroscopic state of the system, as opposed to the instantaneous microscopic state. The work done by the fields is given by the integral of this quantity.

\subsection{Tradeoffs in Control Strategies}
\label{sec:tradeoffs}
Eq.~\ref{eq:PowerDensity22}, along with the dynamics of Eq.~\ref{eq:PhiDyn3}, now allows us to understand dynamics, total thermodynamic cost, and network performance for any control strategy by evaluating a small number of degrees of freedom, which we now use to explore thermodynamically efficient driving in large networks. In this subsection we illustrate this. We use a control
strategy chosen for illustrative purposes only, which is not optimal in sense.

Given a sequence of partial memories, an optimal driving strategy would successfully  complete each memory while minimizing the work done in Eq.~\ref{eq:PowerDensity22} and the time taken $t_f-t_0$. As before, we assume that $\bm{u}(t_0)=\bm{u}(t_f)=0$ and that the system is well localized to a single memory at $t_0$. In this scenario, the entropy produced over the trajectory is simply the work done multiplied by a factor of $\beta$. Note however that the change in free energy at intermediate times  is still nonzero. If the state of the system is not localized to a memory after driving concludes, there is an additional entropy production cost associated with free energy differences. We will focus on the localized  case here.

\begin{figure}
    \centering
    \includegraphics[width=.95\textwidth]{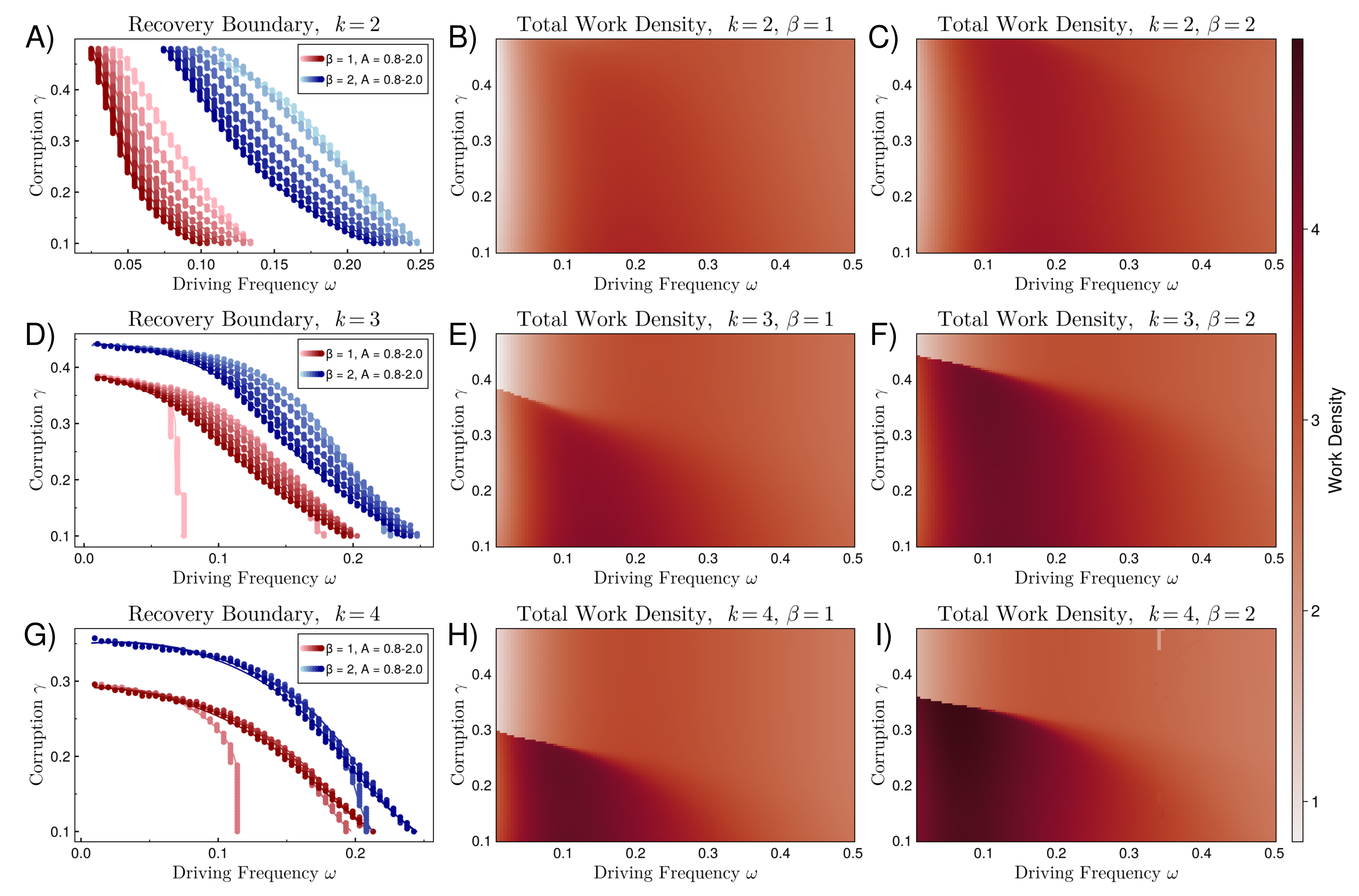}
    \caption{
    Recovery performance and work cost for DenseAM networks. \textbf{(A,D,G)} Recovery boundaries for $k=2,3,4$ memory networks at two temperatures and various driving amplitudes. To the left of the boundaries, each memory in the sequence is recovered within at least 95\% accuracy. At lower temperature, networks can pattern complete more corrupted patterns, and at faster driving there are regimes where  higher order networks are more robust to fast driving at high temperatures. Increasing the driving amplitude increases, then decreases performance, as discussed in the main text.  \textbf{(B,C,E,F,H,I)} The total work density for the \textbf{(B,C)} $k=2$, \textbf{(E,F)} $k=3$, and \textbf{(H,I)} $k=4$ networks at \textbf{(B,E,H)} $\beta=1$ and \textbf{(C,F,I)} $\beta=2$ under the driving strategy in Eq.~\ref{eq:controlStrat2}, for fixed driving amplitude. The higher order networks consume more power under this strategy. In the regime where driving is successful, total work costs typically grow with driving frequency.
    }
    \label{fig:Phases}
\end{figure}

The general control problem is non-convex in the fields, and so we leave the solution to the problem of finding that optimal control protocol to future work. Instead, we characterize control with a small number of parameters of interest, to demonstrate an application of Eq.~\ref{eq:PowerDensity22} in understanding total work costs.  If the network is localized to some pattern, and we want to drive it to a new pattern $\nu_1$ in time interval $[t_0,t_0+\frac{1}{\omega}]$, we consider a family of control strategies of the form:
\begin{align}
    u^\mu(t) = 
    \begin{cases}
    A (1 - \cos(2\pi\omega (t-t_0)))\delta_{\mu,\nu_1} & \text{if } t \in [t_0,t_0+\frac{1}{\omega}] \\
    0, & \text{otherwise}
    \end{cases}
    \label{eq:controlStrat1}
\end{align}
This family is parametrized by $A$, $\omega$, and implicitly by the corruption fraction $\gamma$ and inverse temperature $\beta$, each of which reflects an aspect of network operation. When this control strategy is succesful, it pins the state of the system to the partial memory $\bm{\zeta}^{\nu_1}$ during the first half of the interval, and then allows the network to relax into the actual pattern $\bm{\xi}^{\nu_1}$  as $u$ fades (Fig.~\ref{fig:MFT}). The strategy of Eq.~\ref{eq:controlStrat1} can  be chained together to drive the system through a sequence of  memories, retrieved from a sequence of partial memories $\{\bm{\zeta}^{\nu_1}, \bm{\zeta}^{\nu_2}...\}$. This control strategy is shown in Fig.~\ref{fig:MFT} (C), and  can be formally written as:
\begin{equation}
u^\mu(t) =
\sum_{\nu_l} 
A (1 - \cos(2 \pi \omega  (t - t_{l}))) \, \delta_{\mu,\nu_i} \;\;\; ;
\quad 
t_l =  \frac{l-1}{\omega}
\label{eq:controlStrat2}
\end{equation}
Now given parameters $\omega$, $A$, $\beta$, and $\gamma$, we can  characterize the total power consumption and work done, network operation speed, and the extent of successful memory recovery by simulating  Eqs.~\ref{eq:PhiDyn3} and \ref{eq:PowerDensity22} (Fig.~\ref{fig:Phases}). While the family of control strategies we are considering is not necessarily optimal, we can nevertheless use it compare the thermodynamics of DenseAM networks with various driving regimes and nonlinearities. \textbf{This leads to our last set of results.}

Qualitatively, we  find that memory reconstruction becomes harder at higher driving frequencies, in the sense that larger error rates $\gamma$ must be corrected more slowly (smaller $\omega$) then smaller error rates, for any inverse temperature $\beta$ and driving amplitude $A$ (Fig.~\ref{fig:Phases}). This makes sense: we are considering a  driving strategy that pins the state of the network to partial patterns which then relax into the true memory. The relaxation time without driving  increases with the error fraction $\gamma$ as seen in Fig. ~\ref{fig:dynamics}. Additionally performance does not increase monotonically with the driving amplitude $A$ (Fig.~\ref{fig:Phases}). In fact, the performance increases and then decreases with $A$ for a given driving frequency $\omega$. This decrease in performance as $A$ grows too large is a consequence of the particular class of driving strategies that we are considering, and likely not a fundamental constraint. Indeed, at large $A$ in our procedure, the network remains pinned to partial patterns for longer, and so there is less time time for the network to relax into consecutive memories before the driving field moves on to subsequent partial patterns. As the network will always lag behind the external drive due to its finite response time, excessive  driving can  degrade performance by effectively reducing the time available for successful transitions between patterns. 

Finally, as temperature increases, pattern recovery becomes more difficult, just as in the case without driving. However, we observe from simulations (see Fig.~\ref{fig:Phases}) that there are regimes at intermediate temperature where the higher order networks remain more robust to fast driving then the lower order networks. This is likely due to the higher order networks having steeper basins of attraction, and faster relaxation times at intermediate temperatures, as is explicitly shown in the undriven case in Fig. ~\ref{fig:dynamics}. 

We can compare the thermodynamic cost of different strategies, and find that  higher order networks incur a higher work cost (Fig.~\ref{fig:Phases}) when memory recovery is successful for the class of control strategies considered here. We observe that higher order networks dissipate more energy in the regime of proper memory recall than lower order networks, as shown in Fig.~\ref{fig:Phases}. Although this observation is limited to a restricted family of control strategies, we expect this to hold more generally, and examine the optimally driven case for each network in future work. We expect this trend to persist under broader control strategies, as the energy landscape of higher-order networks is steeper near memory minima but flatter away from them. Heuristically, the steepness associated with higher order networks (greater $k$) forces the system to overcome strong local curvature in the energy landscape, leading to dissipation that is less evenly distributed over the network trajectory, even under optimal control strategies not considered here. As such, we might expect that the higher order networks incur a greater work cost under finite time driving in more generic settings.  We additionally observe that slow driving incurs lower work costs (Fig.~\ref{fig:Phases}), which is a typical feature of thermodynamic systems.

In the adiabatic limit, we expect vanishing dissipation and work cost associated with driving a system between equal free energy minima. Interestingly, work cost  decreases again at fast driving, in the regime where memory recovery fails (Fig.~\ref{fig:Phases}). This can occur for two reasons. First, the system lags the external drive to such an extent that the change in the external fields $\bm{h}$ is usually not aligned with system state, and so the work done per unit time on the network is small, analagous to spinning one's wheels in the mud. This is what causes the decrease in work cost at faster driving in Fig.~\ref{fig:Phases}. The attractor at zero alignment for $k>2$ networks can also contribute to the decline in work at fast driving.
If the network is localized to a pattern A, and then is quickly presented a partial pattern B, the system may instead relax towards the attractor at zero alignment. In this case, presenting the network with partial patterns too quickly will cause the system state to remain within basin boundaries because of the finite response time, and so the state will slide back to the zero basin.
As discussed previously, the work done and entropy produced over the full trajectory are equal up to a factor of $\beta$ for driving strategies that finish with the network localized to a single pattern.

\section{Discussion}
\label{sec:discussion}
In this work, we characterized the dynamics and thermodynamics of DenseAM networks  in a mean field analysis that applies for large networks and below saturation of the memory capacity. Higher order networks of this kind  have a substantially higher memory capacity \cite{patternRecognition}, and so we sought to compare the energetic cost of operating them with that of lower order networks.  We found that when operated via relaxation and at finite temperature, higher order networks sometimes relax away from stored memories  towards a metastable network configuration with vanishing memory alignment. Thus, for any given error rate in the partial memories that they seek to reconstruct, higher order networks must be operated at lower temperatures, leading to greater energy dissipation, and hence entropy production. 
However, when reconstruction is successful, higher order networks also reproduce the target memories with greater accuracy, and are less susceptible to finite temperature statistical fluctuations.

We  explored the energetic cost of actively driving these networks through sequences of corrupted memories. At low memory load $p<<\mathcal{O}(N^{k-1})$ the dynamics can be expressed deterministically in terms of alignment with a small number of memories. As a result,  we can efficiently study the thermodynamic cost of control via numerical simulation.  Using this approach, we examined a  family of control strategies for polynomial DenseAM networks, and  found tradeoffs between the speed, reconstruction accuracy, and thermodynamic cost of memory recall.  In particular, for successful recall we must drive networks more slowly if they are at higher temperature or if the partial memories are more corrupted. At fixed temperature, faster driving additionally incurs higher work cost in regimes where memory reconstruction is successful.  The entropy production in this case equals the work cost times the inverse temperature. We found in general that while higher-order networks have increased storage capacity and better reconstruction accuracy, they incur greater power cost and require stronger control fields.  Conversely, lower-order networks are more thermodynamically efficient at low memory loads, highlighting a fundamental balance between computational capacity and energy efficiency under our choice of control.

We focused on the low memory load regime.  It would be useful to extend our work to understand the thermodynamic cost  of operating DenseAM networks of various orders near saturation of their memory capacity.
At loads close to network capacity, or if the system size $N$ is sufficiently small, stochastic fluctuations become important so that the dynamics of the memory alignments will no longer be well-approximated by the deterministic mean field equations derived here. 
One approach for addressing this challenge might be to write a path integral description of the stochastic dynamics, and use this description to generate  correlation functions in $\phi$ (this is the Martin-Siggia-Rosa-Janssen-De Dominicis (MSRJD) generating functional approach \cite{MSRJD1,MSRJD2,MSRJD3}). For example, by including first order dTAP-like corrections \cite{dTap} to the mean field dynamics described here, one might be able to understand thermodynamic costs for arbitrary load at small to intermediate $N$. Further consideration of the Kramers-Moyal expansion derived in the appendix would also lead naturally to finite $N$ corrections to the mean field dynamics described here at low to intermediate loads.  

To illustrate our methods, we studied a natural family of control strategies. It should be possible to use a similar approach to explore tradeoffs between speed, accuracy, and thermodynamic cost more generally, 
with the goal of finding optimal solutions to the broader network control problem.
For example, this has been demonstrated in small systems such as optimal single particle control \cite{Referee7} and finite time bit erasure \cite{Referee8}. Understanding optimal finite time driving with respect to thermal costs in the coarsening of extremely large networks using the tools of, for example \cite{Referee7,Adrianne1}, is the subject of ongoing work.

It would be especially interesting to compare optimal operation in the low and high memory load regimes, as we expect a qualitative difference: the controller will have to incorporate ongoing adaptive changes to its strategy at high load and finite temperature in order to compensate for the greater effects of stochasticity arising from spin glass degrees of freedom.  Finally, it would be interesting to extend the dynamic mean field analysis that led to our results to study the thermodynamic cost of computation with other, possibly more natural,  neural network architectures. 

\vspace{0.1 in}
\noindent {\bf Acknowledgments: } This work was supported in part by the NSF and DoD OUSD (R \& E) under Agreement PHY-2229929 
(The NSF AI Institute for Artificial and Natural Intelligence).  While this research was in progress, VB was supported in part by the Eastman Professorship at Balliol College, Oxford. Part of this work was done while DK was employed by IBM Research. At the time of submission DK is no longer employed by IBM Research.

\bibliographystyle{abbrvnat}
\bibliography{Refs}

\input{CurrentAppendix}

\end{document}

%% file: CurrentAppendix.tex
\appendix
\section{MFT for Dense Associative Nets}
With no inputs, the Hamiltonian is for the general dense 
(polynomial) associative network of order $k$ is:
\begin{equation}
    \mathcal{H} = -\frac{1}{N^{k-1}}\sum_\mu (\bm{\xi}^\mu\cdot \bm{\sigma})^k
\end{equation}
Here, the normalization keeps the energy density order 1. We insert 
delta functions for each memory alignment/magnetization $\mu$. 
\begin{align}
    \mathcal{Z} &= \sum_{\{\bm\sigma\}} \exp[\frac{\beta}{N^{k-1}}\sum_\mu (\bm\xi^\mu \cdot \bm\sigma)^k] \\
    &= C \, \sum_{\{\bm\sigma\}} \int \prod_\mu d\phi^\mu  \, 
                \delta(\phi^\mu- \frac{1}{N}\bm\xi^\mu \cdot \bm\sigma)
                \exp[N\beta \sum_\mu (\bm\phi^\mu)^k]
    \\ 
    & = C \, \sum_{\{\bm\sigma\}} \int D[\phi^\mu,\tilde{\phi}^\mu] \,  e^{N\sum_\mu 
                \tilde{\phi}^\mu (\phi^\mu-\frac{1}{N}\bm\xi^\mu \cdot \bm\sigma)+N\beta \sum_\mu (\bm\phi^\mu)^k)}       
\end{align}
Here $C$ absorbs overall constant factors that play no role in the analysis, and $D[]$ is  shorthand for the integral measure.  The memory alignment variables lie in the range $-1 \leq \phi^\mu \leq 1$, and integrating over them with inserted delta functions sets $\phi^\mu = \frac{1}{N}\bm\xi^\mu \cdot \bm\sigma$, thus reproducing the explicit partition function.  In the second line we have used a standard representation over the delta function where we integrate over complex conjugate fields $\tilde\phi$ along a contour on the imaginary axis from $-i\infty$ to $+i\infty$. The spins are now decoupled, and we can perform the sum on $\{\bm\sigma\}$ as before:
\begin{align}
    \sum_{\{\bm\sigma\}}e^{-\sum_\mu \tilde{\phi}^\mu \bm\xi^\mu\cdot\bm\sigma}
    &=
    \prod_i 2\cosh(\sum_\mu \tilde{\phi}^\mu \xi^\mu_i)  
=  2^N \exp[\sum_i^N \ln\cosh(\sum_\mu \tilde{\phi}^\mu \xi^\mu_i)]\\
\Longrightarrow \ \ 
\mathcal{Z} &= 
        C \int D[\phi^\mu,\tilde{\phi}^\mu] \,  e^{-N\mathcal{S}[\bm\phi,\bm \tilde{\bm\phi};\{\xi^\mu\}]} \\ 
       \mathcal{S} &= - \sum_\mu \tilde{\phi}^\mu\phi^\mu - {1 \over N}
        \sum_i^N \ln\cosh(\sum_\mu \tilde{\phi}^\mu \xi^\mu_i) -
        \beta \sum_\mu (\phi^\mu)^k
        \label{eq:DANEffectiveAction}
\end{align}
where once again we introduced an {\it effective action} $\mathcal{S}$ and absorbed the factor of $2^N$ from the first line into the normalization constant $C$.

In the large $N$ limit we expect the partition function to be dominated by the saddlepoints of the effective action. The saddlepoint values of $\phi^\mu$ are then {\it mean fields} representing the average alignment of the spins with with the memory $\xi^\mu$ in the configuration that dominates the partition function.  Recall that the $\tilde{\phi}^\mu$ integrals above run along the imaginary axis, and so $\mathcal{S}$ can be complex. Following the method of steepest descent \cite{book2} for approximating  complex integrals, we should deform the integration contour to run through stationary points of the integrand such that the real part of $-\mathcal{S}$ is  concave down in every argument along the contour of integration thus giving a local maximum, while the imaginary part is constant in the vicinity of the saddle thus locally eliminating oscillations. At large $N$ the partition sum will be well approximated by the sum of values evaluated at stationary points that lie on such contours of steepest descent.  The steepest descent stationary points in $\tilde{\phi}$ need not lie on the imaginary axis along which the integral was originally defined. MFT becomes exact in the sense that the following limit 
\begin{align}
    \lim_{N\rightarrow\infty} \frac{1}{N}\ln\int D[\phi^\mu,\tilde{\phi}^\mu] \,  e^{-N\mathcal{S}[\bm\phi,\bm \tilde{\bm\phi};\{\xi^\mu\}]} 
\end{align}
approaches the effective action evaluated at the saddles.

\subsubsection{Single Memory}
To establish the  procedure, we start with the one memory case:
\begin{align}
    \mathcal{S}[\phi,\tilde{\phi};\bm\xi] &= - \tilde{\phi}\phi -
        \frac{1}{N}\sum_i^N \ln\cosh(\tilde{\phi}\xi_i) -
        \beta \phi^k\\
        &=
        - \tilde{\phi}\phi -
        \ln\cosh(\tilde{\phi}) -
        \beta \phi^k
\end{align}
where $\phi$ is real and lies between $-1$ and $1$.   The initial choice of contour for $\tilde{\phi}$ in the partition function integral takes $\tilde{\phi}$ along the imaginary axis. 
But a steepest descent contour passing through a stationary point  $\tilde{\phi}^*$  of the integral need not lie on the imaginary line \cite{book1,book2}.  Indeed, we can smoothly deform the contour to pass through $\tilde{\phi}^*$ so long as we do not  pass through poles of the integrrand. Fortunately, $\mathcal{S}$ has no poles in $\tilde{\phi}$, though it has logarithmic branch points at $\tilde{\phi}=i(\frac{\pi}{2}+\pi\mathbb{Z})$. Furthermore, there is always a choice of contour passing through $\tilde{\phi}^*$ such that $\Im[S]$ is constant in
the neighbourhood of $\tilde{\phi}^*$ and along the contour \cite{book1}.
Now, we know that the partition sum we started with and the  free energy are real; this means that $\Im[S]$ in the neighbourhood of the saddle must also vanish. As $\phi$ is also real by definition, this means that at $\tilde{\phi}^*$, either $\tilde{\phi}$ is real or $\Im[\tilde{\phi}]=-\Im[\frac{1}{\phi N}\sum_i^N \ln\cosh(\tilde{\phi}\xi_i)]$. We will find that at the saddle point $\tilde\phi$ is real.

Explicitly, we extremize the effective action by finding where variations $\delta S$ vanish. Letting $\tilde\varphi$ and $\tilde{\psi}$ be the real and complex parts of $\tilde\phi$ respectively, $\tilde\phi=\tilde\varphi+i\tilde{\psi}$, this amounts to requiring that:
\begin{align}
    \delta\mathcal{S} = \frac{\partial S}{\partial \phi}\delta\phi + \frac{\partial S}{\partial \tilde{\varphi}}\delta{\tilde\varphi} + \frac{\partial S}{\partial \tilde\psi}\delta{\tilde\psi} = 0
\end{align}
Setting each derivative to zero yields:
\begin{align}
    \frac{\partial}{\partial \phi}\mathcal{S} &= -\tilde{\phi}-k\beta \phi^{k-1}=0  \\
    \frac{\partial}{\partial \tilde\varphi}\mathcal{S}
    &=-\phi-\frac{1}{N}\sum_j\tanh{(\tilde\varphi+i\tilde\psi)\xi_j}=0\\
    \frac{\partial}{\partial \tilde\psi}\mathcal{S}
    &=-i\phi-\frac{i}{N}\sum_j\tanh{[(\tilde\varphi+i\tilde\psi)\xi_j]}=0   
\end{align}
Note  that the last two conditions are the same. This is a result of the fact that for functions $g$ whose complex derivative exists,  the derivative $g'(z)$ is independent of the angle of approach in the complex plane, and variations $g(z+\delta z)-g(z) = g'(z)\delta z =g'(z) \delta r e^{i\theta}$ are equivalent up to the angle of the variation. In shorthand, we write: 
\begin{align}
    \frac{\partial}{\partial \phi}\mathcal{S} &= -\tilde{\phi}-k\beta \phi^{k-1}=0 \label{eq:DANsaddle1} \\
    \frac{\partial}{\partial \tilde{\phi}}\mathcal{S} &
    = -\phi-\frac{1}{N}\sum_i\xi_i\tanh(\tilde{\phi}\xi_i)=-\phi-\tanh(\tilde{\phi})=0  \label{eq:DANsaddle2}
\end{align}
where in the second line we used the facts that $\xi_i = \pm 1$ and that  $\tanh$ is an odd function of its arguments.  We can use the first equation to eliminate $\tilde\phi$ in the second equation, to arrive at a self consistency condition for $\phi$:
\begin{align}
    \phi = \tanh(k\beta \phi^{k-1})  \label{eq:DANSelfConsistent}
\end{align}
This equation always has a solution at $\phi=0$ representing no alignment. When the temperature is very low (large $\beta$) the $\tanh$ has a sharp slope at $\phi = 0$ and saturates to a value of $\pm1$, so there are also solutions for $\phi \sim \pm 1$ representing almost perfect alignment or anti-alignment with the memory. There will be a critical value of the temperature above which these aligned solutions vanish, meaning that the memory cannot be recovered as an equilibrium configuration.  Likewise, the solution at $\phi=0$ remains stable unless $k=2$, in which case it is unstable at sufficiently low temperature, so that the only solutions are aligned with the stored memory, as described in the main text. In the one memory case, the free energy can be written purely in terms of $\phi$ by inserting $\tilde{\phi}^\star$ back into the effective action:
\begin{align}
    \partial_{\tilde{\phi}} \mathcal{S}[\phi,\tilde{\phi}]|_{\tilde{\phi}^*} &=0\rightarrow \tilde{\phi}^*=-\arctanh(\phi)\\
    \tilde{\mathcal{S}}[\phi] &= \mathcal{S}[\phi,\tilde{\phi}]|_{\tilde{\phi}^*}
    =\phi\arctanh(\phi) -\ln\cosh(\arctanh(\phi))-\beta\phi^k\\
    &=-\beta\phi^k + \frac{1}{2}[(1-\phi)\ln(1-\phi)+(1+\phi)\ln(1+\phi)]
\end{align}

\subsubsection{Multiple Memories}
For DenseAM networks storing $p$ memories, we should extremize the effective action (\ref{eq:DANEffectiveAction}).
Extremizing with respect to $\tilde{\phi}^\mu$ gives
\begin{equation}
\tilde{\phi}^{\mu *} = -\beta k (\phi^{\mu *})^{k-1} 
\label{eq:k-consistency1}
\end{equation}
Next, extremizing (\ref{eq:DANEffectiveAction}) with respect to $\tilde{\phi}^\mu$ and insert the solution (\ref{eq:k-consistency1}) for $\tilde{\phi}^\mu$ into the resulting equation.  This gives

\begin{align}
    \phi^{\mu *} &= 
    \frac{1}{N}\sum_i \xi_i^\mu \tanh(k\beta\sum_\nu (\phi^{\nu*})^{k-1}\xi_i^\nu) 
    \label{eq:selfConsistMain}  
    \\
    &=
    \frac{1}{N}\sum_i  \tanh(k\beta\sum_\nu (\phi^{\nu*})^{k-1}\xi_i^\nu\xi_i^\mu) 
    \\
    &=
    \frac{1}{N}\sum_i  \tanh\Big(k\beta \, \Big[(\phi^{\mu*})^{k-1}+\sum_{\nu\neq\mu} (\phi^{\nu*})^{k-1}\xi^\mu\xi^\nu \Big]\Big) 
    \\
    &=
    \mathbb{E}_{\bm{x}^\nu} \left[\tanh\Big(k\beta \, \Big[(\phi^{\mu*})^{k-1}+\sum_{\nu\neq\mu} (\phi^{\nu*})^{k-1}x^\nu \Big]\Big) \right] \, .
\label{eq:k-consistency2}
\end{align}
where $x^\nu = \pm 1$ have equal probability. For $k=2$, (\ref{eq:k-consistency2}) agrees with the self consistency equation for the quadratic model.  Unlike the one memory case, setting the gradient of $\mathcal{S}$ with respect to $\bm{\tilde\phi}$ to zero leads to an equation that is not uniquely invertible, and so it is not possible to express the free energy in terms of $\bm{\phi}$ alone away from fixed points.  At fixed points, the free energy density can be found by inserting Eq.~\ref{eq:selfConsistMain} back into the effective action. 

If a state $\bm\sigma$ remains  correlated with a memory $\bm\xi^\mu$ in the large $N$ limit, then the corresponding alignment $\phi^\mu = \frac{1}{N} \sum_{i=1}^N \xi^\mu_i \sigma_i$ will be $O(1)$ since $\sigma_i =\pm 1$ and $\xi^\mu_i = \pm 1$ will tend to have the same sign and there are $N$ terms in the sum.  If the state is uncorrelated with a memory then the sign of $\sigma_i \xi^\mu_i$ will be $\pm 1$ with equal probability.   So, the expected value of these $\phi^\mu$ will be zero, with a standard deviation of $O(1/\sqrt{N})$. Suppose for a particular solution of the self-consistency conditions, $S = \{\phi^{\mu_1},...\phi^{\mu_a}\}$ is the set of $O(1)$ alignments and $S_{na}$ is the set of $O(1/\sqrt{N})$ alignments.  
Then, we can split the sum in the self-consistency equation  (\ref{eq:k-consistency2}) as
\begin{align}
    \phi^{\mu*}=\mathbb{E}_{\bm{x}} 
    \left[
    \tanh \Big(
    k\beta \Big[
   (\phi^{\mu*})^{k-1}+
    \sum_{\nu\in S; \nu\neq\mu} (\phi^{\nu*})^{k-1}x^{\nu}
    +\sum_{\kappa\in S_{na}; \kappa \neq \mu}(\phi^{\kappa*})^{k-1}x^{\kappa} \Big]
    \Big)  
    \right]
    \label{eq:FinalEqSelf}
\end{align}
We want to look for solutions in which $a$, the number of memories with which the state is aligned is $O(1)$ and  much smaller than $p$, the number of stored memories. If $\mu$ indexes an aligned memory, that the first two sums in (\ref{eq:FinalEqSelf}) are $O(1)$. Now consider the last sum which contains the contribution of the unaligned memories.  Each term in the sum is independently distributed with zero mean and has standard deviation $1/N^{(k-1)/2}$, while $x^\kappa = \pm 1$ with equal probability. So the sum will have zero mean, and standard deviation of $O(\sqrt{p}/N^{(k-1)/2})$ where $p\gg a$ is the total number of stored memories. So for this last term to compete with the first two the network must be storing $p \sim O(N^{k-1})$ memories.  For smaller loads,  unaligned patterns will not contribute to the mean field self-consistency condition, and hence  to the equilibrium free energy. In other words, away from saturation of the memory capacity one can use the self-consistency condition (\ref{eq:selfConsistMain}) applied to the $O(1)$ alignments instead of all $O(p)$ terms to understand the equilibrium free energy of the system.  Explicitly, the sum over the noncondensed patterns is a gaussian random variable by the central limit theoerm \cite{Sompolinsky2,StudentTeacher}:
\begin{align}
    \sum_{\kappa\in \mathcal{S}_{na}}(\phi^{\kappa})^{k-1}x^{\kappa}\rightarrow
    z\sim\mathcal{N}(0,\sigma^2_{\phi})
\end{align}
whose variance is vanishing in the low to intermediate load regimes, but nonvanishing near saturation.

\section{The Kramers-Moyal Expansion}

We argued heuristically that the dynamics in each $\phi$ are non-stochastic and approach their mean if  $p<<\mathcal{O}(N^{k-1})$.
We demonstrate that this statement is exact when $p\sim\mathcal{O}(\sqrt{N})$ via the Kramers-Moyal expansion \cite{Kramers,Moyal,Pawula,coolen}.
Similar arguments have been used to show that the $k=2$ (Hopfield) dynamics are deterministic in their alignments at sufficiently low load \cite{coolen}.

We start by writing the probability distribution in the macroscopic degrees of freedom via:
\begin{align}
    P(\bm\Phi) = \sum_{\{\bm{\sigma}\}} P(\bm{\sigma},t)\delta{(\bm\Phi-\bm\Phi(\bm{\sigma})}) 
\end{align}
Here $\bm\Phi(\bm\sigma)$ is shorthand for $\bm\Phi(\bm\sigma) = \frac{1}{N}[\bm\xi^1 \cdot \bm{\sigma},\bm\xi^2 \cdot \bm{\sigma},...]$, whereas the delta function enforces this definition of $\bm\Phi$, so that the delta function is shorthand for the product of delta functions:
\begin{align}
    \delta{(\bm\Phi-\bm\Phi(\bm{\sigma}))} \equiv \prod_\mu \delta(\phi^\mu-\frac{1}{N}\bm\xi^\mu \cdot \bm{\sigma})
\end{align}
Taking a time derivative on both sides yields:
\begin{align}
    \partial_t P(\bm\Phi,t) &= \sum_{\{\bm{\sigma}\}} \frac{d}{dt}P(\bm{\sigma},t)\delta{(\bm\Phi-\bm\Phi(\bm{\sigma})})\\
    &=\sum_{\bm\sigma} \sum_i \left[ \Gamma_i(S_i\bm{\sigma};t)P(S_i\bm{\sigma};t) 
    -\Gamma_i(\bm{\sigma};t)P(\bm{\sigma};t) \right]\delta{(\bm\Phi-\bm\Phi(\bm{\sigma})})
    \label{eq:Kramers-Moyal1}
\end{align}
Where we have used the master equation in $\bm\sigma$ (eq.~\ref{eq:Master1}) in the second line.

To proceed, we will utilize the following identity:
\begin{align}
    \delta(\bm y-\bm x-\bm\varepsilon) &= e^{-\bm \varepsilon \cdot \nabla_{\bm y}} \delta(\bm y- \bm x)
    \label{eq:deltaId}
\end{align}

Derivatives of the delta function (and distributions in general) should be interpreted by their action on test functions via integration by parts, under which the above identity follows naturally from representations of the above in Fourier space.

The dynamics in the probability distribution of $\bm\Phi$
can be split into two terms:
\begin{align}
    \partial_t P(\bm\Phi,t) 
    =&  \sum_{\bm\sigma} \sum_i \left[
    \Gamma_i(S_i\bm{\sigma};t)P(S_i\bm{\sigma};t)
    \delta{(\bm\Phi-\bm\Phi(\bm{\sigma})})
    \right]\\
    -&\sum_{\bm\sigma} \sum_i \left[
    \Gamma_i(\bm{\sigma};t)P(\bm{\sigma};t)
    \delta{(\bm\Phi-\bm\Phi(\bm{\sigma})})\right] 
    \label{eq:Kramer-Moyal2}
\end{align}
We focus on the first term:
\begin{align}
    &\sum_{\bm\sigma} \sum_i \left[ 
    \Gamma_i(S_i\bm{\sigma};t)P(S_i\bm{\sigma};t)
    \delta{(\bm\Phi-\bm\Phi(\bm{\sigma})})\right]
    \label{eq:First1}
    \\
    =&
    \sum_{\bm\sigma} \sum_i \left[ 
    \Gamma_i(\bm{\sigma};t)P(\bm{\sigma};t)
    \delta{(\bm\Phi-\bm\Phi(\bm{S_i\sigma})})\right]
    \label{eq:First2}\\
    =&
    \sum_{\bm\sigma} \sum_i \left[ 
    \Gamma_i(\bm{\sigma};t)P(\bm{\sigma};t)
    \delta{(\bm\Phi-\bm\Phi(\bm{\sigma})}-\bm\Delta_i(\bm{\sigma}))\right]
    \label{eq:First3}\\
    =&
    \sum_{\bm\sigma} \sum_i \left[ 
    \Gamma_i(\bm{\sigma};t)P(\bm{\sigma};t)e^{-\bm\Delta_i(\bm{\sigma})\cdot \nabla_{\bm\Phi}}
    \delta{(\bm\Phi-\bm\Phi(\bm{\sigma})})\right]
    \label{eq:First4}
\end{align}
In going from equation~\ref{eq:First1} to equation~\ref{eq:First2}, we have taken advantage of the fact that the trace is invariant under local gauge transformations $\sigma_i\rightarrow-\sigma_i$.
To go from equation~\ref{eq:First2} to equation~\ref{eq:First3}, we have defined the vector $\bm\Delta_i$ as the difference between $\bm\Phi(\bm{S_i\sigma})$ and $\bm\Phi(\bm{\sigma})$, whose entries are $\Delta_i^\mu$.
Concretely:
\begin{align}
    \bm\Delta_i(\bm\sigma) &\equiv \bm\Phi(\bm{S_i\sigma})-\bm\Phi(\bm{\sigma})\\
    \Delta_i^\mu(\bm{\sigma}) &\equiv \phi^\mu(\bm{S_i\sigma})-\phi^\mu(\bm{\sigma}) = -\frac{2}{N}\xi^\mu_i\sigma_i
\end{align}
Finally, we have used the delta function identity of equation~\ref{eq:deltaId} to go from equation~\ref{eq:First3} to equation~\ref{eq:First4}.
Plugging this back into equation~\ref{eq:Kramer-Moyal2} yields:
\begin{align}
    \partial_t P(\bm\Phi,t) 
    =&  \sum_{\bm\sigma} \sum_i \left[ 
    \Gamma_i(\bm{\sigma};t)P(\bm{\sigma};t)e^{-\bm\Delta_i\cdot \nabla_{\bm\Phi}}
    \delta{(\bm\Phi-\bm\Phi(\bm{\sigma})})\right]\\
    -&\sum_{\bm\sigma} \sum_i \left[
    \Gamma_i(\bm{\sigma};t)P(\bm{\sigma};t)
    \delta{(\bm\Phi-\bm\Phi(\bm{\sigma})})\right]\\
    \rightarrow
    \partial_t P(\bm\Phi,t)
    =&\sum_{\bm\sigma} \sum_i \left[ 
    \Gamma_i(\bm{\sigma};t)P(\bm{\sigma};t)[e^{-\sum_\mu\Delta_i^\mu \partial_{\mu}}-1]
    \delta{(\bm\Phi-\bm\Phi(\bm{\sigma})})\right]
    \label{eq:Kramer-Moyal3}
\end{align}
Where $\partial_\mu\equiv \frac{\partial}{\partial \phi^\mu}$. 
We can now expand the exponential. 
As each $\Delta_i^\mu$ is small, $\Delta_i^\mu=-\frac{2}{N}\xi^\mu_i\sigma_i$, this expansion can be truncated to first order if $p\sim\mathcal{O}(\sqrt{N})$.
For loads greater than this, such a truncation requires more thought, as although each $\Delta_i^\mu$ is small in $N$, the collective effect from summing on $i$ and $\mu$ may compensate.
We expand the exponential, and keep the quadratic term for illustrative purposes:
\begin{align}
    e^{-\sum_\mu\Delta_i^\mu \partial_{\mu}}-1 = 
    -\sum_\mu\Delta_i^\mu \partial_{\mu}
    +\frac{1}{2}\sum_\mu\sum_\nu \Delta_i^\mu\Delta_i^\nu \partial_{\mu}\partial_{\nu}-\mathcal{O}(\Delta^3)
\end{align}
Plugging this back into equation~\ref{eq:Kramer-Moyal3} and rearranging, one has:
\begin{align}
    \partial_t P(\bm\Phi,t) &= -\sum_\mu \partial_\mu J^\mu(\bm{\Phi},t)+ \frac{1}{2}\sum_{\mu,\nu}\partial_\mu\partial_\nu K^{\mu\nu}(\bm\Phi,t)
\end{align}
Where we have defined:
\begin{align}
    J^\mu &\equiv \sum_{\bm\sigma}\sum_i\left[ 
    \Gamma_i(\bm{\sigma};t)P(\bm{\sigma};t)\Delta_i^\mu
    \delta{(\bm\Phi-\bm\Phi(\bm{\sigma})})\right]
    \\
    K^{\mu\nu} &\equiv \sum_{\bm\sigma}\sum_i\left[ 
    \Gamma_i(\bm{\sigma};t)P(\bm{\sigma};t)\Delta_i^\mu\Delta_i^\nu
    \delta{(\bm\Phi-\bm\Phi(\bm{\sigma})})\right]
\end{align}
We now must handle these two terms. 
To proceed, it is convenient to consider the probability of microscopic state $\bm{\sigma}$, conditioned on macroscopic state $\bm\Phi$:
By Baye's rule,
\begin{align}
    P(\bm\sigma|\bm\Phi;t) = \frac{P(\bm\Phi|\bm\sigma;t)P(\bm{\sigma};t)}{P(\bm\Phi;t)}
    =\frac{\delta(\bm\Phi-\bm\Phi(\bm\sigma))P(\bm{\sigma};t)}{P(\bm\Phi;t)}
\end{align}
If we let $\langle . \rangle_{\bm\Phi}$ represent an expectation with respect to this distribution, 
$\langle . \rangle_{\bm\Phi} = \sum_{\bm\sigma} . P(\bm\sigma|\bm\Phi;t) $, then both $J^\mu$ and $K^{\mu\nu}$ can be expressed in terms of these expectations. 
Starting with $J^\mu$, we have:
\begin{align}
    J^\mu &= \bigg\langle \sum_i \Gamma_i(\bm{\sigma},t)\Delta_i^\mu \bigg\rangle_{\bm\Phi}
    P(\bm\Phi,t)\\
    &=\bigg\langle \sum_i \frac{1}{2}(1-\tanh[k\beta \sum_\nu (\phi^\nu(\bm\sigma))^{k-1}\xi_i^\nu\sigma_i])(-2\frac{1}{N}\sigma_i\xi_i^\mu) \bigg\rangle_{\bm\Phi}
    P(\bm\Phi,t)\\
    &=\frac{1}{N}\sum_i \bigg\langle  -\xi_i^\mu\sigma_i+\tanh[k\beta \sum_\nu (\phi^\nu(\bm\sigma))^{k-1}\xi_i^\nu\xi_i^\mu] \bigg\rangle_{\bm\Phi}
    P(\bm\Phi,t)\\
    &=\Bigg[-\phi^\mu  + 
    \mathbb{E}_{\bm{x}}\tanh\Big[ k\beta(\phi^\mu)^{k-1} + k\beta\sum_{\nu\neq\mu}  (\phi^\nu)^{k-1}x^\nu\Big]\Bigg]P(\bm\Phi,t)
    \label{eq:AMu}
    \\
    &\equiv A^\mu(\bm{\Phi},t) P(\bm\Phi,t)
\end{align}
Where we have defined $A^\mu$ via the prefactor of equation~\ref{eq:AMu}. 
Likewise, for $K^{\mu\nu}$ one has:
\begin{align}
    K^{\mu\nu}&= \bigg\langle \sum_i \Gamma_i(\bm{\sigma},t)\Delta_i^\mu\Delta_i^\nu \bigg\rangle_{\bm\Phi}
    P(\bm\Phi,t)\\
    &=\bigg\langle \sum_i \frac{1}{2}(1-\tanh[k\beta \sum_\eta(\phi^\eta(\bm\sigma))^{k-1}\xi_i^\eta\sigma_i])(4\frac{1}{N^2}\xi_i^\nu\xi_i^\mu) \bigg\rangle_{\bm\Phi}P(\bm\Phi,t)\\
    &=\frac{2}{N^2}\sum_i(\xi_i^\nu\xi_i^\mu)(1-\langle\sigma_i\rangle_{\bm\Phi}\tanh[k\beta \sum_\eta(\phi^\eta(\bm\sigma))^{k-1}\xi_i^\eta])P(\bm\Phi,t)\\
    &\equiv B^{\mu\nu}(\bm{\Phi},t) P(\bm\Phi,t)
\end{align}
Each $\xi_i^\nu\xi_i^\mu$ term is one on the diagonal, and a $\pm 1$ RV where $\mu\neq \nu$.
The scalar part is bounded in $[0,2]$, so the diagonal elements can, at worst, scale as $\frac{1}{N}$, with off diagonal elements scaling as $\frac{1}{N^{3/2}}$.

Putting it all together, the master equation in $P(\bm\Phi)$ can be written in the form of a Fokker-Plank equation:
\begin{align}
    \partial_t P(\bm\Phi,t) &= 
    -\sum_\mu \partial_\mu[A^\mu(\bm{\Phi},t)P(\bm\Phi,t)] 
    + \frac{1}{2}\sum_{\mu\nu}\partial_\mu\partial_\nu[B^{\mu\nu}(\bm{\Phi},t) P(\bm\Phi,t)]
\end{align}

I.e., dynamics in $\bm\Phi$ yield a drift-diffusion (Brownian) process. 

Although we kept the quadratic term in the expansion of the exponential, the effect of the diffusion term $B^{\mu\nu}$ is vanishing if $p$ scales as $\mathcal{O}(\sqrt{N})$ or slower.
In this regime, the dynamics are deterministic and given by $\partial_t \phi^\mu = A^\mu(\phi)$, which matches the relaxation dynamics of the last section. 
At finite $N$, the diffusion term provides the leading order stochastic correction to the dynamics arising from finite size effects.
Heuristically, this is in line with the finite $N$ trajectories shown in Fig.~\ref{fig:dynamics}, which appear Brownian about the deterministic drift process.

At loads $p>\sqrt{N}$, truncating the exponential in equation~\ref{eq:Kramer-Moyal3} requires further arguments about the scale of the action of the derivatives $\partial_\mu$ on the probability distribution at short timescales. 
Alternatively, one can split the dynamics in the alignments into two sectors $\bm\Phi = [\bm\Phi^{U_c},\bm\Phi^{U_{nc}}]$, where $\bm{\Phi^{U_{nc}}}$ is the set of patterns $\phi^\mu$ which remain unaligned over a short time window. 
Marginalizing the distribution in $P(\bm\Phi)$ over the $\bm\Phi^{U_{nc}}$ unaligned patterns before expanding the exponential, when possible, might lead to a valid truncation.
We pursue neither approach here, instead arguing heuristically for the scaling given in the main text.

\section{Numerical Details}
All numerics were performed using Julia \cite{Julia}, partially through the use of the DifferentialEquations.jl \cite{DL_jl} package. Visualizations were created with CairoMakie \cite{Makie}.